\documentclass[aps,prb,twocolumn,reprint]{revtex4-1}
\usepackage{amsmath,amssymb,bm,braket,bbm}
\usepackage[pdftex]{graphicx}
\usepackage{color}
\usepackage{comment}
\usepackage{braket}
\usepackage{mathtools}

\usepackage{docmute}

\usepackage{mathptmx}
\usepackage{accents}
\newcommand\ddd[2]{\accentset{\circ}{#1}_{#2}}

\newcommand\dddLL[2]{\accentset{\circ}{#1}^{\, \mathrm{L}}_{#2}}
\newcommand\dddRR[2]{\accentset{\circ}{#1}^{\, \mathrm{R}}_{#2}}
\newcommand\drst[2]{\accentset{\circ}{#1}^{\, \prime}_{#2}}
\def\drk#1{\drst{k}{#1}}
\def\dre#1{\drst{e}{#1}}

\newcommand{\imi}{\mathrm{i}}
\newcommand{\clam}{\Lambda}
\newcommand{\diff}{\mathrm{d}}
\newcommand{\paren}[1]{\left(#1\right)}
\newcommand{\bck}[1]{\left[#1\right]}
\newcommand{\bce}[1]{\left\{#1\right\}}
\newcommand{\asub}[2]{\left. #1\right|_{#2}}

\DeclareMathOperator{\sech}{sech}

\def\LL{{\mathrm{L}}}
\def\RR{{\mathrm{R}}}

\begin{document}
\title{
Generalized hydrodynamic approach to charge 
and energy currents in the one-dimensional Hubbard model}

\author{Yuji Nozawa and Hirokazu Tsunetsugu}
\affiliation{
The Institute for Solid State Physics, 
The University of Tokyo, Kashiwanoha 5-1-5, Chiba 277-8581, Japan}

\date{\today}
\begin{abstract}
We have studied nonequilibrium dynamics of the one-dimensional 
Hubbard model using the generalized hydrodynamic theory.
We mainly investigated the spatio-temporal profile of 
particle density (equivalent to charge density), energy density 
and their currents using the partitioning protocol; 
the initial state consists of two semi-infinite different 
thermal equilibrium states joined at the origin.
In this protocol, there appears around the origin a transient region 
where currents flow, and this region expands its size linearly 
with time.  
We examined how density and current profiles depend on initial 
conditions. 
Inside the transient region, 
we have found a \textit{clogged region} where charge current 
is zero but nonvanishing energy current flows.
This phenomenon is similar to spin-charge separation in Tomonaga-Luttinger liquids, 
in which spin and charge excitations propagate with different velocities.
This region appears when one of the initial states has 
half-filled electron density, 
and it is located adjacent to this initial state.  
We have proved analytically the existence of the clogged region 
in the infinite temperature case of the half-filled initial state.  
The existence is confirmed also for finite temperatures 
by numerical calculations of generalized hydrodynamics.
A similar analytical proof is also given for a clogged region of 
spin current when magnetic field is applied to one and 
only one of the two initial states.  
A universal proportionality of charge and spin currents 
is also proved for a special region, for general initial 
conditions of electron density and magnetic field.  
To examine the clogged region of charge current, we have 
calculated the current contributions of different types of 
quasiparticles in the Hubbard model. 
It is found that the charge current component carried 
by scattering states (i.e. Fermi liquid type) is canceled 
completely in the clogged region 
by the counter flows carried by bound-state quasiparticles 
(called $k$-$\clam$ strings).  
Their contributions are not canceled for energy current, 
which is also proved analytically for special cases, 
and these different behaviors are the origin of the clogged region.  
Except for the clogged region, charge and energy densities 
are in good proportion to each other, and their ratio 
depends on the initial conditions.  
This proportionality in nonequilibrium dynamics is 
reminiscent of Wiedemann-Franz law in thermal equilibrium, 
which states current proportionality determined by temperature.  
The long-time 
stationary 
values of charge and energy currents 
were also studied with varying initial conditions.  
We have compared the results with the values of non-interacting 
systems and discussed the effects of electron correlations.
The ratio of these stationary currents is also analyzed.
We have found that the temperature dependence of the ratio
is strongly suppressed by electron correlations and
even reversed.
\end{abstract}
\maketitle

\section{Introduction}
\label{sec:1}
One-dimensional (1D) integrable models play special roles in the studies of nonequilibrium dynamics of quantum many-body systems. Their eigenstates can be obtained through the Bethe ansatz method~\cite{bethe1931theorie}. Integrable models have an extensive number of conserved quantities~\cite{korepin_bogoliubov_izergin_1993}. Their existence leads to anomalous transport properties such as nonzero Drude weights at finite temperatures~\cite{PhysRevLett.74.972,PhysRevB.55.11029}, 
where the Drude weight is an important quantity in the linear response theory~\cite{PhysRev.133.A171}. In addition, these conserved quantities constrain the equilibration of the system. It is conjectured that the long-time asymptotic stationary state is described by the generalized Gibbs ensemble~\cite{PhysRevLett.98.050405} in integrable models.

The generalized hydrodynamics (GHD) was developed recently~\cite{PhysRevX.6.041065,PhysRevLett.117.207201}
to describe dynamics in integrable models. The time evolution equations of GHD are derived from the continuity equations of conserved quantities and represented in terms of the distribution functions of quasiparticles in these models.
In particular, the partitioning protocol~\cite{rubin1971abnormal,spohn1977stationary,bernard2012energy,Bernard2015,bhaseen2015energy,PhysRevX.6.041065,PhysRevLett.117.207201} has been under intensive study 
because it is simpler than the other protocols in GHD. 
In this protocol, two equilibrium states are connected
at time $t=0$, and its time evolution is studied as we will explain in Sec.~\ref{sec:3}. By using this protocol, various transport properties, for example, time dependence of currents~\cite{PhysRevX.6.041065,PhysRevLett.117.207201,fagotti2016charges,PhysRevB.96.020403,doyon2017dynamics,PhysRevB.97.045407,PhysRevLett.120.045301,PhysRevB.97.081111,Bertini_2018,PhysRevLett.120.176801,10.21468/SciPostPhys.4.6.045,PhysRevB.98.075421,PhysRevB.99.014305,PhysRevB.99.174203,doi:10.1063/1.5096892,Bulchandani_2019}, Drude weights~\cite{PhysRevLett.119.020602,PhysRevB.96.081118,SciPostPhys.3.6.039}, entanglements~\cite{PhysRevB.97.245135,Bertini_ent,PhysRevB.99.045150,10.21468/SciPostPhys.7.1.005}, correlation functions of densities and currents~\cite{PhysRevB.96.115124,10.21468/SciPostPhys.5.5.054}, diffusive dynamics and diffusion constants~\cite{PhysRevLett.121.160603,10.21468/SciPostPhys.6.4.049,PhysRevLett.121.230602} have been studied.
It should be noted that the validity of GHD was recently confirmed by an experiment in a 1D Bose gas system on an atom chip~\cite{PhysRevLett.122.090601}.

The 1D Hubbard model, which is studied in this paper, is a standard lattice model of 
strongly correlated electrons and exactly
solved by the nested Bethe ansatz~\cite{PhysRevLett.19.1312, GAUDIN196755, PhysRevLett.21.192.2,essler2005one}. 
The model has two degrees of freedom, charge and spin, and correspondingly the types of quasiparticles in the model are classified into scattering and bound states of 
 charge and spin. Ilievski and De Nardis applied GHD to the 1D Hubbard model for the first time in Ref.~29~\nocite{PhysRevB.96.081118} and mainly studied its Drude weight. They also used the partitioning protocol and compared their results with the numerical results in Ref.~45\nocite{PhysRevB.95.115148}. 
However, the dependence of currents on initial conditions was not studied systematically.

In this paper, we examine systematically the dependence of charge and energy currents on chemical potential and temperature 
in an initial state in the partitioning protocol. 
We have found a spatial region that has no charge current while energy current flows if one part of the initial state is half-filled, i.e. one electron per site. We study this region analytically in some limiting cases and numerically in more general cases. We examine the contributions of different types of quasiparticles to charge and energy currents 
and show that different contributions cancel to each other in the charge current in the region. A similar phenomenon was found in the XXZ model, and no spin current flows under some conditions~\cite{PhysRevB.96.115124}. Our result is analogous to this  regarding charge degrees of freedom. We also show the presence of a similar phenomenon regarding spin current. In addition, we discuss general relationships between charge and spin currents.

We also study stationary charge and energy currents as time goes to infinity. We analyze systematically the dependence of the stationary currents on the initial conditions. We show that the initial temperature dependence changes qualitatively with the initial particle density. We investigate the effects of the Coulomb interaction on the stationary charge current, 
by comparing with non-interacting cases. 

This paper is organized as follows. In Sec.~\ref{sec:2}, the model and its Bethe ansatz formulation are described. In Sec.~\ref{sec:3}, we review the protocol considered in this paper and the generalized hydrodynamic theory. In Sec.~\ref{sec:4}, we examine analytically the profiles of local densities and their currents in some limiting cases. In Sec.~\ref{sec:5}, we consider numerically these quantities at finite temperatures.
The conclusion is given in Sec.~\ref{sec:6}.

\section{The one-dimensional Hubbard model}
\label{sec:2}
The Hamiltonian of the 1D Hubbard model on $L$ sites is given by 
\begin{align}
\hat{H}&
=-\sum_{j=1}^{L} \sum_{\sigma} 
\left[
\bigl( \hat{c}_{j,\sigma}^{\dagger}\hat{c}_{j+1,\sigma}+\mathrm{H.c.} \bigr)
+ (\mu + s_{\sigma} B ) \hat{n}_{j, \sigma}
\right]
\nonumber\\
&+4u \sum_{j=1}^{L}
\bck{ 
\bigl( \hat{n}_{j,\uparrow}-{\textstyle \frac{1}{2}} \bigr)
\bigl( \hat{n}_{j,\downarrow}-{\textstyle \frac{1}{2}} \bigr)-{\textstyle \frac{1}{4}}},
\label{eq:Hamiltonian}
\end{align}
where $\hat{c}_{j,\sigma}^{\dagger}$ and $\hat{c}_{j,\sigma}$ 
are the electron creation and annihilation operator, respectively, 
at site $j$ with spin 
$\sigma\  \in \{\uparrow,\downarrow \}$. 
$\hat{n}_{j,\sigma} \equiv 
\hat{c}^{\dagger}_{j,\sigma} \hat{c}_{j,\sigma}$ and 
$s_{\sigma}$ is defined as $s_{\uparrow}=1$ and $s_{\downarrow}=-1$.
We have set the electron hopping amplitude to be unity, 
and use it as the unit of energy throughout this paper.
$\mu$ and $B$ are chemical potential and magnetic field, respectively.  
The Coulomb repulsion is parameterized by $u~ >0$, 
and the constant $-1/4$ in this term is included so as to
make the energy of the vacuum state zero. 

A remarkable point of the 1D Hubbard model 
is that it is exactly solvable by the nested Bethe ansatz~\cite{PhysRevLett.19.1312, GAUDIN196755, PhysRevLett.21.192.2,essler2005one}. 
Each eigenstate of the Hamiltonian with $N_{\uparrow}$ spin-up electrons and $N_{\downarrow}$ spin-down electrons is described 
by a set of charge momenta $\bm{k}=(k_{1}, \cdots, k_{N_{\uparrow}+N_{\downarrow}})$ and spin rapidities $\bm{\lambda}=(\lambda_{1},\cdots,\lambda_{N_{\downarrow}})$ 
, which are determined by solving the Lieb-Wu equations~\cite{PhysRevLett.21.192.2}. 

\begin{figure}[tb]
\centering
\includegraphics[width=\columnwidth,clip]{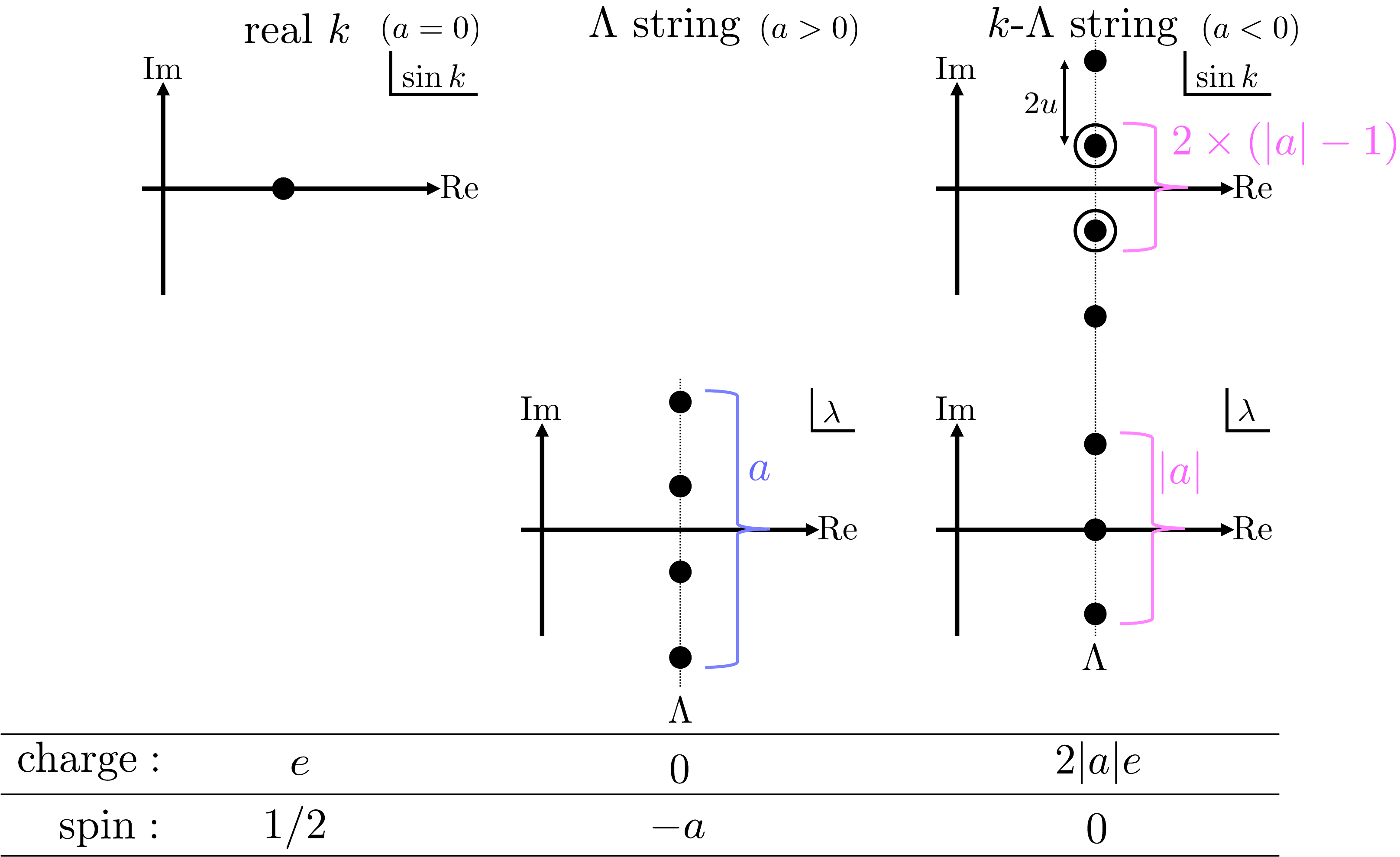}
\caption{
Three types of roots of the Lieb-Wu equations: 
(Left) real $k$, (Center) $\clam$ string, and (Right) $k$-$\clam$ string.  
Examples of $\clam$ string for $a=4$ and $k$-$\clam$ string for $a=-3$ 
are shown.
Black dots in
each complex plane
represent roots. 
Those with circle are doubly degenerate corresponding to $k$ and $\pi-k$, 
which have the same value of $\sin k$.  
These three are different types of quasiparticles in the Hubbard model, 
and their charge and spin are listed at the bottom.
$e(<0)$ is the electron charge.
}
\label{fig:string}
\end{figure}

In the thermodynamic limit, the string hypothesis~\cite{10.1143/PTP.47.69} claims 
that the positions of $k$'s and $\lambda$'s in the complex plane 
are described by three types of distribution functions  
\begin{equation}
\rho_0 (k), \quad
\{\rho_a (\clam ) \}_{a=1,2,\cdots}, \quad
\{ \rho_a (\clam ) \}_{a=-1,-2,\cdots}.  
\end{equation}
The first one $(a=0)$ is the distribution of real $k$, 
and the second one $(a>0)$ is about $\clam$ string 
made of $a$ complex $\lambda$'s equally separated along the imaginary 
axis with the common real part $\clam$.  
The last one $(a<0 )$ is about $k$-$\clam$ string 
made of equally separated $\lambda$'s together with $\sin k$'s 
at the common real part $\clam$.  
The number of $\lambda$'s and $k$'s is $|a|$ and $2|a|$ respectively.
The configurations of $k$'s and $\lambda$'s in these types and their charge and spin are shown in Fig.~\ref{fig:string}.
As will be shown in Eq.~\eqref{eq:ev}, in GHD, the time evolution is described by the dynamics of distribution functions of each configuration in Fig.~\ref{fig:string} labeled by an integer $a$. 
Therefore, we call it a quasiparticle of type-$a$. Real $k$'s $(a=0)$ are scattering states and each of them carries charge $e$ and spin $1/2$. Each $\clam$ string $a(>0)$ carries spin $a$. Each $k$-$\clam$ string $a(<0)$ is 
a bound state~\cite{essler2005one} carrying charge $2|a|e$ and spin $0$.

For describing a thermal equilibrium state, 
one also needs their hole distributions $\rho_a^h$.  
Instead of the pair of $\rho_a$ and $\rho_a^h$, 
one can alternatively use the total distribution 
$\rho_a^t$ and the filling function $\vartheta_a$
\begin{equation}
\rho_a^t \equiv \rho_a + \rho_a^h , \quad
\vartheta_a \equiv \frac{\rho_a }{ \rho_a^t }.  
\end{equation}
For later use, we also define different parametrizations
\begin{equation}
\bar{\vartheta}_a \equiv \frac{ \rho_a^h }{ \rho_a^t } = 1 - \vartheta_a , 
\quad
\eta_a \equiv \frac{\rho_a^h}{\rho_a} = \vartheta_a^{-1} - 1 . 
\label{eq2:eta}
\end{equation}

\section{Generalized hydrodynamic approach and partitioning protocol}
\label{sec:3}
\begin{figure}[tb]
\centering
\includegraphics[width=\columnwidth,clip]{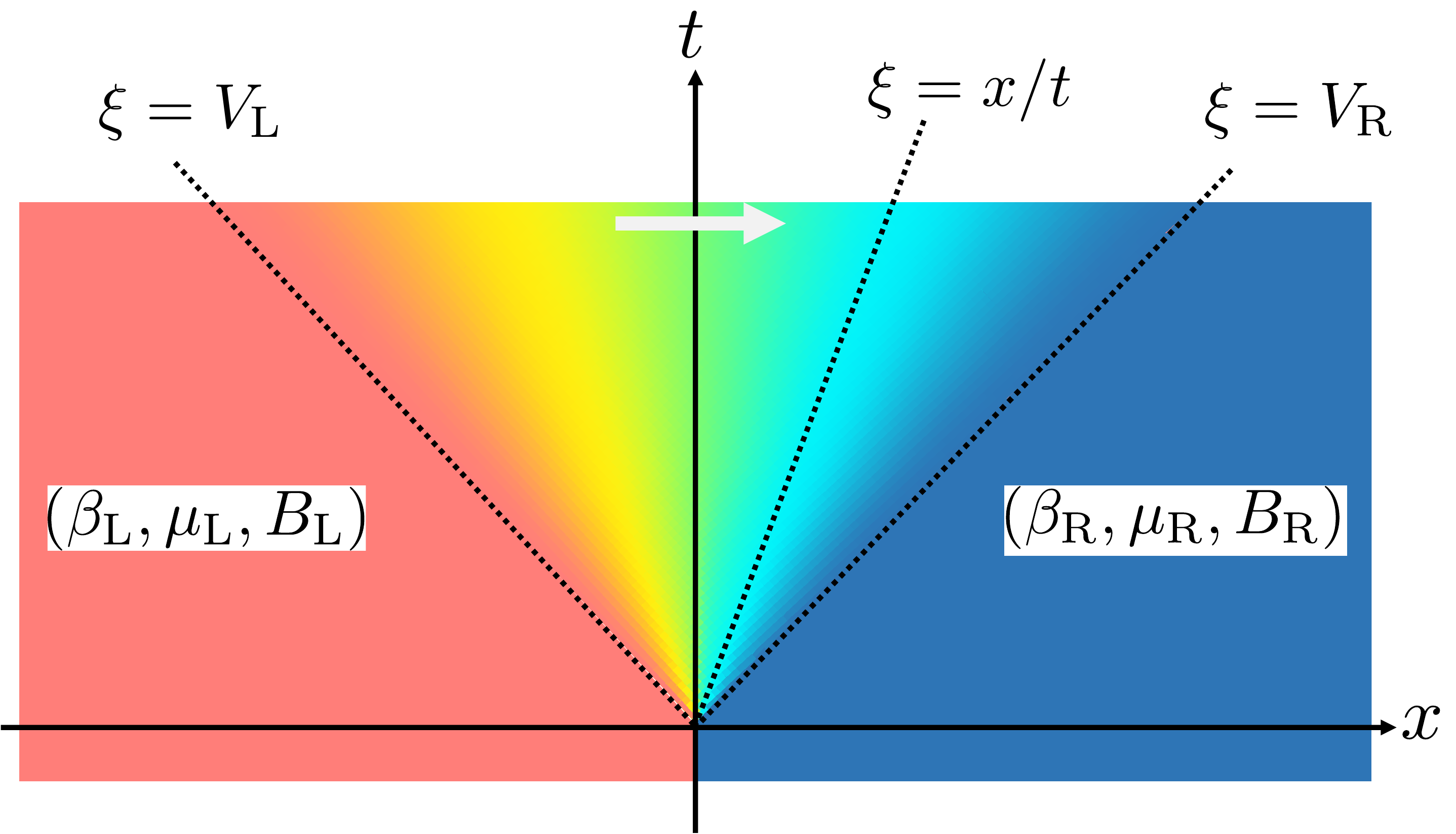}
\caption{
Schematic picture of partitioning protocol. 
The initial state consists of two thermal equilibrium states; 
they are disconnected until $t=0$ and joined at the origin $x=0$ after that.
The left and right states are prepared with the control parameters 
such as inverse temperature $\beta_{\LL(\RR)}$, 
chemical potential $\mu_{\LL(\RR)}$, and magnetic field $B_{\LL(\RR)}$. 
Local distributions of quasiparticles at $(x, t)$ 
depend only on the ray $\xi=x/t$. 
Outside the light cones $\xi\leq V_{\LL}$ or $\xi \geq V_{\RR}$, the distributions are unchanged from those in the initial left or right thermal equilibrium state, respectively.
}
\label{fig:setup}
\end{figure}

We analyze the nonequilibrium dynamics using the partitioning protocol. In this protocol, two semi-infinite chains are kept 
in different equilibria until time $t=0$ as shown in Fig.~\ref{fig:setup}. The initial left (right) thermal equilibrium state is determined by the inverse temperature $\beta_{\LL(\RR)}$, the chemical potential $\mu_{\LL(\RR)}$, and the magnetic field $B_{\LL(\RR)}$. $V_{\LL(\RR)}$ is the left (right) light cone, and if the ray $\xi=x/t$ satisfies 
$\xi\leq V_{\LL}$ or $\xi \geq V_{\RR}$, the local state remains the initial left (right) equilibrium state.
In this paper, we consider the case that $\mu_{\mathrm{s}}\leq 0$ and $B_{\mathrm{s}}\geq 0$ for $\mathrm{s}=\mathrm{L},\mathrm{R}$, which means the initial left (right) particle density $n^{\LL (\RR)}$ and magnetization $m^{\LL (\RR)}$ satisfy $n^{\LL (\RR)}\leq 1$ and $m^{\LL (\RR)}\geq 0$, respectively.

In this section, we outline the formulation 
for the partitioning protocol in the 1D Hubbard model following Ref.~29\nocite{PhysRevB.96.081118}.
Since the GHD describes the large-scale dynamics, we use a coarse-grained continuous variable $x$ instead of $j$.
In GHD, the quantum state of the integrable system is represented 
by the distributions at each space-time point 
$\{ \rho_a (w; x,t) , \rho_a^h (w; x,t) \}$, 
and the time evolution follows 
the continuity equations of quasiparticles 
corresponding to real-$k$'s and each type of strings
\begin{align}
  \frac{\partial}{\partial t} \, \rho_{a}\paren{w;x,t} 
+ \frac{\partial}{\partial x} \, 
  \left[ \ddd{v}{a} (w;x,t) \rho_{a} (w;x,t) \right] 
=0, 
\label{eq:ev}
\end{align}
where $w = k$ for $a=0$ or otherwise $w=\clam$.  
$\{ \ddd{v}{a} \}$ are the dressed velocities~\cite{PhysRevLett.113.187203},
which will be explained later. 
In the following, quantities with a small circle atop 
denote their dressed values.

A picture that particles with charge and spin degrees of freedom propagate separately with different velocities 
has been well-known for 1D electron systems. Their low temperature properties are described by bosonization and Tomonaga-Luttinger liquid theory~\cite{Haldane_1981,PhysRevLett.47.1840,doi:10.1143/JPSJ.58.3752,PhysRevB.41.2326,PhysRevLett.64.2831,Kawakami_1991,PhysRevB.42.10553,giamarchi2003quantum,gogolin2004bosonization}, and this phenomenon is known as spin-charge separation~\cite{PhysRevLett.77.4054,Segobia_1999,PhysRevLett.90.020401,PhysRevLett.95.176401,PhysRevLett.98.266403,PhysRevA.77.013607,jompol2009probing,PhysRevB.82.245104,PhysRevLett.104.116403,PhysRevB.85.085414}, where excitations with charge and spin are called holons and spinons, respectively. A characteristic point of GHD is that this theory is not restricted to low temperature regimes but covers the whole temperature range up to infinity, although it is applicable only to integrable systems. Additionally, the number of types of quasiparticles is not a small number but infinite. It reflects the presence of infinite number of conserved quantities corresponding to string solutions of Bethe ansatz equations. Using GHD, spin-charge separation effects were studied in the Yang-Gaudin model~\cite{PhysRevB.99.014305}, which is the continuum limit of the 1D Hubbard model~\cite{essler2005one}.

\def\ssp{4pt}

In the partitioning protocol, the distributions and physical 
observables depend only on the ray 
\begin{equation}
 \xi = \frac{x}{t}, \quad (t > 0) . 
\end{equation}
Quantities under consideration are a local density 
of conserved quantity and its corresponding 
current density such as particle density $n (\xi)$ 
and its current $j_n (\xi)$.
Note that charge current $j_c$ is proportional to the particle 
density current
\begin{equation}
  j_c (\xi ) = e \, j_n (\xi ),
\end{equation}
where $e < 0$ is the electron charge, and therefore
essentially they are identical.
Henceforth in this paper, we will calculate $j_n$ and discuss 
charge current based on its data.
Readers should be warned that the two words, particle density current and
charge current, are used interchangeably throughout this paper.  
They can be calculated from the quasiparticle distributions 
$\{ \rho_a (w \maketitle ,\xi )\}$ 
and their dressed velocities $ \ddd{v}{a} (w , \xi )$.  
The particle density $n$, 
magnetization $m$, and energy density $e$ 
and their currents $j_n, j_m, j_e $ are given by
\begin{align}
\left[ 
\begin{array}{c}
n (\xi ) \\[\ssp ] 
j_n (\xi )
\end{array}
\right]
&= 
\left[ 
\begin{array}{c}
\tilde{n}_0 (\xi ) \\[\ssp ]  
\tilde{j}_0 (\xi ) 
\end{array}
\right]
+ \sum_{a<0} 2 |a| \, 
\left[ 
\begin{array}{c}
\tilde{n}_a (\xi ) \\[\ssp ]  
\tilde{j}_a (\xi ) 
\end{array}
\right],
\label{eq:n}
\\
\left[ 
\begin{array}{c}
m (\xi ) \\[\ssp ]  
j_m (\xi )
\end{array}
\right]
&= {\textstyle \frac{1}{2}} 
\left[ 
\begin{array}{c}
\tilde{n}_0 (\xi ) \\[\ssp ]  
\tilde{j}_0 (\xi ) 
\end{array}
\right]
- \sum_{a>0} a \, 
\left[ 
\begin{array}{c}
\tilde{n}_a (\xi ) \\[\ssp ]  
\tilde{j}_a (\xi ) 
\end{array}
\right],
\label{eq:m}
\\
\left[ 
\begin{array}{c}
e (\xi ) \\[\ssp ]  
j_e (\xi )
\end{array}
\right]
&= 
\sum_{a \le 0} 
\left[ 
\begin{array}{c}
\tilde{e}_a (\xi ) \\[\ssp ]  
\tilde{\kappa}_{a} (\xi ) 
\end{array}
\right],
\label{eq:e}
\end{align}
with 
\begin{align}
\left[
\begin{array}{c}
\displaystyle
\tilde{n}_a (\xi ) \\[\ssp ]  
\displaystyle
\tilde{e}_a (\xi ) \\[\ssp ]  
\displaystyle
\tilde{j}_a (\xi ) \\[\ssp ]  
\displaystyle
\tilde{\kappa}_a (\xi ) 
\end{array}
\right]
&\equiv  \int dw \, 
\left[
\begin{array}{c}
\displaystyle
1  \\[\ssp ]  
\displaystyle
e_a (w ) \\[\ssp ]  
\displaystyle
\ddd{v}{a} (w , \xi ) \\[\ssp ]  
\displaystyle
e_a (w ) \ddd{v}{a} (w , \xi )
\end{array}
\right]
\rho_a (w ,\xi ). 
\label{eq3:density_component}
\end{align}
Here, $e_{a}$ is the bare energy of the type-$a$ quasiparticle:  
\begin{equation}
 e_{0} (k)= -2 \cos k  - 2u , 
\quad
 e_{a < 0} (\clam ) = 
4\Re \sqrt{1-(\clam + i a u)^2}
+4au, 
\label{eq:be}
\end{equation}
and $e _{a>0} (\clam ) = 0$.
The symbol $\Re$ denotes the real part.  
By using Takahashi's equations~\eqref{eqD:te1}-\eqref{eqD:te4}, 
alternative expressions are obtained for $n$ and $m$
\begin{align}
n (\xi ) 
&=
1- \int_{-\infty}^{\infty} \! d \clam \, \rho^{t}_{-\infty} (\clam , \xi ) 
\label{eq:n2}
\\
m (\xi ) 
&= {\textstyle \frac{1}{2} } 
\int_{-\infty}^{\infty} \! d \clam \, \rho^{t}_{\infty} (\clam , \xi ), 
\label{eq:m2}
\end{align}
and these will be used afterwards. An equation analogous to Eq.~\eqref{eq:m2} also holds in the XXZ model~\cite{PhysRevB.96.115124}. 

Now, let us sketch how to obtain $\{ \rho_a (w,\xi ) \}$ 
and $\{ \ddd{v}{a} (w , \xi ) \}$.  
As will be shown below, it is easier to calculate 
the filling function $\{ \vartheta_a (w,\xi ) \}$ 
together with $\{ \ddd{v}{a} (w , \xi ) \}$. 
In the process of solving them, 
the total distributions $\{ \rho^{t}_a (w,\xi ) \}$ 
are obtained and the quasiparticle distributions 
are determined as $\rho_a = \vartheta_a \, \rho_a^t$.  

\begin{figure}[tb]
\centering
\includegraphics[width=6.0cm,clip]{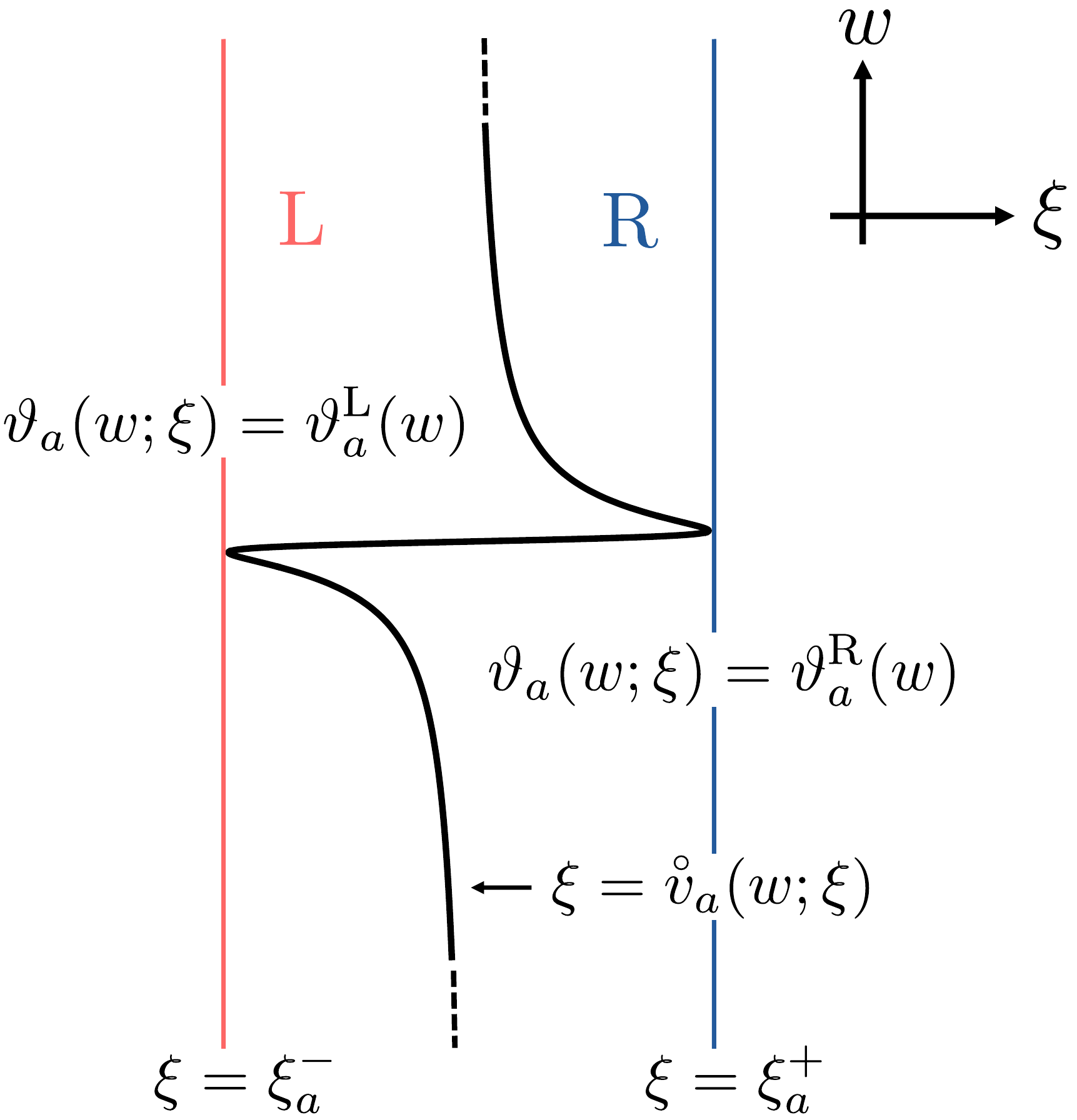}
\caption{
Sketch of how to determine the values of the filling function 
$\vartheta_{a}(w,\xi)$. 
The curve $\xi=\overset{\circ}{v}_{a} (w,\xi)$ divides 
the $\xi$-$w$ space into two parts, 
and the values of the light cones $\xi^{-}_{a}$ and $\xi^{+}_{a}$ 
are determined as the leftmost and rightmost points, respectively, 
on the curve. 
In the left (right) area, the filling function is determined 
as $\vartheta_{a}(w,\xi)=\vartheta^{\LL(\RR)}_{a}(w)$.
}
\label{fig:fillfig}
\end{figure}

The two quantities, 
$\{ \ddd{v}{a} \}$ and $\{ \vartheta_a \}$ are 
related to each other. 
Therefore, one should calculate them consistently, and 
we use iterations for that.  
Suppose the dressed velocities are given and 
we try to calculate the filling functions $\{ \vartheta_a (w,\xi ) \}$.   
They follow the differential equation~\cite{PhysRevB.96.081118} 
\begin{align}
  \frac{\partial}{\partial t} \, \vartheta_{a} (w;x,t)
+ \ddd{v}{a} (w;x,t) \frac{\partial}{\partial x} \, 
 \vartheta_{a} (w;x,t) 
=0, 
\label{eq:ev2}
\end{align} 
and its initial condition is 
\begin{equation}
 \vartheta_{a} (w;x,t=0) 
= \Theta (-x) \vartheta^{\LL} (w) 
+ \Theta (x)  \vartheta^{\RR} (w) , 
\end{equation}
where $\Theta (x)$ is Heaviside's step function, 
$\Theta (x) =1$ for $x>0$ and 0 otherwise.  
This equation is easier to solve than Eq.~\eqref{eq:ev}, 
because the differentiation does not operate to the velocity.  
The ray representation of this equation reads
\begin{align}
\bigl[ \xi  -  \ddd{v}{a} ( w ,  \xi ) \bigr]  
\frac{\partial}{\partial \xi } \vartheta_{a} (w , \xi ) 
= 0 . 
\end{align}
This means 
$\partial_{\xi} \vartheta_a (w , \xi ) \propto \delta 
( \xi  -  \ddd{v}{a} ( w ,  \xi ) )$, and 
integrating this leads to the solution of the filling function
\begin{equation}
 \vartheta_a (w, \xi ) 
= 
   \Theta \left( \ddd{v}{a} ( w ,  \xi ) - \xi \right) 
          \vartheta_a^{\LL} (w)
+ \Theta \left( \xi - \ddd{v}{a} ( w ,  \xi ) \right) 
          \vartheta_a^{\RR} (w) . 
\label{eq:vartheta}
\end{equation}
For each string $a$, the curve $ \xi=\ddd{v}{a} ( w ,  \xi ) $ 
divides the $\xi$-$w$ space into two parts 
as shown in Fig.~\ref{fig:fillfig}. 
The filling function is given by $\vartheta_a^{\LL}$ 
in the left part, and by $\vartheta_a^{\RR}$ in the right part. We define the light cones for each string $a$ as 
shown in Fig.~\ref{fig:fillfig}
\begin{align}
\xi_{a}^{-}\equiv \min_{w}\bck{\xi,\xi=\ddd{v}{a} ( w ,  \xi )},\quad \xi_{a}^{+}\equiv \max_{w}\bck{\xi,\xi=\ddd{v}{a} ( w ,  \xi )}.
\end{align}
From Eq.~\eqref{eq:vartheta}, the filling function $\vartheta_{a}(w,\xi)$ satisfies 
\begin{align}
\vartheta_{a}(w,\xi)=\vartheta_a^{\LL}(w),\quad (\xi<\xi_{a}^{-}),\nonumber\\
\vartheta_{a}(w,\xi)=\vartheta_a^{\RR}(w),\quad (\xi>\xi_{a}^{+}).
\end{align}

Thus, the GHD calculation consists of two parts. 
The first part is the calculation of the filling functions 
$\{ \vartheta_a^{\LL} \}$ and 
$\{ \vartheta_a^{\RR} \}$ for the initial equilibrium 
in the two parts.  
It is sufficient to calculate them once 
at the beginning of the whole procedure, 
and the details are explained in Appendix~\ref{app:filling}. 

The second part is about the dressed velocities 
$\{ \ddd{v}{a} \}$, and they should be consistent with 
the filling functions obtained from Eq.~\eqref{eq:vartheta} 
as we will explain below.  
For each $\xi$, 
the dressed velocity is defined in terms of dressed momentum 
$\ddd{k}{a}$ and dressed energy $\ddd{e}{a}$ as 
\begin{align}
  \ddd{v}{a} (w, \xi )
 =\frac{ d  \ddd{e}{a} (w, \xi )}
       { d  \ddd{k}{a} (w, \xi )}
 =\frac{ \dre{a} (w, \xi )}
       { \drk{a} (w, \xi )} , 
\end{align}
where the prime symbol represents the differentiation by $w$. 
These derivatives 
$\{ \drk{a} \}$ and 
$\{ \dre{a} \}$ 
are calculated from the filling functions $\{ \vartheta_a \}$, 
and this part is explained in Appendix~\ref{app:dr}.  
Thus, we need to determine the dressed velocities 
and the filling functions consistently, and we do this by iteration.  
Starting from an appropriate initial candidate 
of $\{ \vartheta_a \}$, we calculate $\{ \drk{a} \}$ and 
$\{ \dre{a} \}$ to obtain $\{ \ddd{v}{a} \}$. 
Using the obtained dressed velocities, we update 
the filling functions using Eq.~\eqref{eq:vartheta}.  
This is one cycle, and we repeat this cycle 
until $\{ \ddd{v}{a} \}$ and $\{ \vartheta_a \}$ 
both converge: 
\begin{equation}
 \{ \vartheta_a \} 
 \rightarrow 
\begin{array}{c}
\{ \drk{a} \} \\[6pt] 
\{ \dre{a} \}
\end{array}
 \rightarrow 
 \{ \ddd{v}{a} \} 
 \rightarrow 
 \{ \vartheta_a \} 
 \rightarrow 
\begin{array}{c}
\{ \drk{a} \} \\[6pt] 
\{ \dre{a} \}
\end{array}
 \rightarrow 
 \cdots.
\end{equation}
The total distributions are immediately obtained from 
the converged result 
\begin{equation}
 \rho_a^t (w , \xi ) = \pm (2\pi )^{-1} 
 \drk{a} (w, \xi ) , 
\end{equation}
where the sign is $+$ for $a \ge 0$ and $-$ for $a<0$. 
These lead to the quasiparticle distributions 
as $\rho_a (w,\xi )$=$\vartheta_a (w,\xi ) \rho_a^t (w , \xi)$. 
Physical quantities are calculated from them 
using Eqs.~\eqref{eq:n}-\eqref{eq3:density_component}.

\section{Analytical results of $n(\xi)$ and $j_{n}(\xi)$ in the high temperature limit}
\label{sec:4}

When the initial equilibrium state is at infinite temperature on one side, 
one can analytically analyze particle density $n(\xi )$, 
magnetization $m(\xi )$, and energy density $e(\xi )$
as well as their current profiles.
We examine the contributions of different quasiparticles 
to those quantities.  
The result shows the presence of a \textit{clogged region}, 
where the charge current is zero but a nonvanishing 
energy current flows.  
It is essential for this region that quasiparticles propagate with different velocities.
The presence of multiple velocities has been known in correlated 1D metals, and spin and charge degrees of freedom propagate with different velocities, which is called spin-charge separation~\cite{PhysRevLett.77.4054,Segobia_1999,PhysRevLett.90.020401,PhysRevLett.95.176401,PhysRevLett.98.266403,PhysRevA.77.013607,jompol2009probing,PhysRevB.82.245104,PhysRevLett.104.116403,PhysRevB.85.085414}. 
This region exists irrespective of the initial equilibrium 
on the other side.  
If the initial temperature is not infinite but high, 
the clogged behavior remains, 
and we will show in the next section 
numerical GHD calculations for the initial state at finite temperatures.  

Let us consider the setup that the initial left equilibrium is at infinite temperature
$\beta_\LL = 0$.  
One can still control the particle density and 
the magnetization by setting nonvanishing parameters 
\begin{equation}
\bar{\mu} \equiv \beta_{\LL} \mu_\LL \le 0 , 
\quad
\bar{B} \equiv \beta_{\LL} B_\LL \ge 0 .  
\end{equation}
The initial state in the right part is arbitrarily set 
by the parameters $\beta_\RR$, $\mu_\RR$, and $B_\RR $.  

For the infinite-temperature initial state in the left part, 
analytic solutions have been known for the thermodynamic Bethe Ansatz 
(TBA) equations 
\eqref{eq:tba4}-\eqref{eq:tba3} 
and Takahashi's equations \eqref{eqD:te1}-\eqref{eqD:te4}~\cite{10.1143/PTP.47.69}. 
Using these solutions, 
we obtained dressed velocities in the initial left part 
\begin{align}
\dddLL{v}{0} (k)
&=
\frac{ \displaystyle
2f_{1} f_{-1} \sin{k} - e^{\prime}_{-2} (\sin k ) \cos{k} }
{ \displaystyle 
 f_{1} f_{-1}         - k^{\prime}_{-2} (\sin k ) \cos{k}  },
\label{eq:infdv1}
\\[6pt]
\dddLL{v}{\pm a} (\clam )
&=
\frac{
  f_{\pm (a+1)} e^{\prime}_{-a}     (\clam ) 
- f_{\pm (a-1)} e^{\prime}_{-(a+2)} (\clam )
}
{
  f_{\pm (a+1)} k^{\prime}_{-a}     (\clam )
- f_{\pm (a-1)} k^{\prime}_{-(a+2)} (\clam )
},
\quad \paren{a>0}.
\label{eq:infdv2}
\end{align}
Here, $k^{\prime}_{a<0}$ is the derivative of the charge momentum of the type-$a$ quasiparticle:
\begin{align}
k^{\prime}_{a}(\clam)=-2\Re\bck{1-\paren{\clam+\imi a u}^{2}}^{-1/2}.
\end{align}
Another simple point is that 
the filling functions are constant in the left part independent 
of $k$ or $\Lambda$
\begin{equation}
\vartheta^{\LL}_{a} =
\begin{dcases}
\paren{1+\frac{\cosh \bar{\mu} }{\cosh \bar{B} }}^{-1}, 
&
\quad \paren{a=0},
\\
f_{a}^{\, -2},
&
\quad \paren{a\neq 0}.
\end{dcases}
\label{eq:inffill}
\end{equation}
These results are represented by the factors
\begin{align}
f_{a} \equiv 
\frac{\sinh [ (|a|+1) \gamma ]}{\sinh \gamma} , 
\quad 
\gamma &=  
\begin{dcases}
\bar{\mu}, 
&
\paren{a < 0},
\\
\bar{B}, 
&
\paren{a>0}, 
\end{dcases}
\label{eq:fa}
\end{align}
and $f_0 =1$ is also defined for later use.  
Note that they are finite even when 
$\bar{\mu}=0$ or $\bar{B}=0$, and 
$f_{a} = |a|+1$ for $a<0$ or $a>0$, respectively.  

From Eqs.~\eqref{eq:infdv2}, 
we have proved that
\begin{align}
    -2< \dddLL{v}{a} (\clam)  ,\quad 
({}^{\forall} a\neq 0) ,
\label{eq:dvine}
\end{align}
for any values of $\bar{\mu}$ and $\bar{B}$.  
The proof is given in Appendix \ref{app:dv}. 
The leftmost light cone has the slope
\begin{equation}
 V_{\LL} = \min_{a} \, \xi_a^{-} 
= \xi_0^{-}  \le -2 . 
\label{eq:leftslope}
\end{equation}
This is because $\ddd{v}{0} ( k=-\pi/2 ) =- 2$. 
For $\xi\leq V_{\LL}$, all of the filling functions $\bce{\vartheta_{a}}$ are 
unchanged from the initial left filling functions $\bce{\vartheta^{\LL}_{a}}$, and therefore 
the local state remains the initial left equilibrium state. 
Let us define similarly $V_{\RR} = \max_a \xi_a^{+}$.  $V_{\LL}$ and $V_{\RR}$ correspond 
to the Lieb-Robinson bounds~\cite{lieb1972}.

For $\xi > V_{\LL} $, the filling functions 
$\bce{\vartheta_{a}} $ generally differ 
from $\bce{\vartheta^{\LL}_{a}}$, and 
hence densities and currents depend 
on the parameters of the initial right equilibrium 
$\beta_{\RR }$, $\mu_{\RR }$, and $B_{\RR }$. 
However, for a specific $\xi$ region, 
we can derive quasiparticles' contributions to densities 
which hold regardless of the right equilibrium. 
In particular, in the case that the initial 
left part is half-filled $\bar{\mu}=0$, we can prove the presence of 
a clogged $\xi$ region where no charge current flows inside 
the light cone. Remember that $\beta_{\LL}=0$.

The clogged state appears in the region 
\begin{equation}
V_{\LL} < \xi < \xi_{-\infty}^{-}. 
\end{equation}
For any $\xi$ in this region, 
there exists a negative integer $a_{*}$ such that 
$\xi < \xi_{a}^{-}$, 
${}^\forall a < a_{*} $.  
Long $k$-$\Lambda$ strings have 
the occupation
\begin{align}
    \vartheta_{a}\paren{\clam,\xi}=\vartheta^{\LL }_{a},
    \quad {}^\forall a<a_{*},
\label{eq4:filling}
\end{align}
and the total distribution has 
a Fourier transformation given by 
\begin{align}
\tilde{\rho}^{t}_{a} \paren{p,\xi} 
&\equiv
\int_{-\infty}^{\infty} \diff\clam \, e^{-i \clam p} \, 
\rho^{t}_{a} (\clam,\xi ) 
\nonumber \\ 
&= A_{1} \paren{p,\xi} f_{a}
 \left(
   \frac{ e^{ a   u|p|} }{ f_{a+1} } 
 - \frac{ e^{(a-2)u|p|} }{ f_{a-1} }
\right) , \quad ( a < a_* ).
\label{eq4:rhotp}
\end{align}
The proof is given in Appendix \ref{app:rhotp}.  
The coefficient $A_{1} \paren{p,\xi}$ is to be determined 
by Takahashi's equations \eqref{eqD:te1}-\eqref{eqD:te4} and depends on the initial state in the right part.
We define the contribution of the type-$a$ quasiparticles to the particle density $n_{a}\paren{\xi}$ as
\begin{align}
n_{0}(\xi)&\equiv\tilde{n}_{0}(\xi),\nonumber\\
n_{a}(\xi)&\equiv2|a|\tilde{n}_{a}(\xi),\quad (a<0).
\end{align}
$n_{a}\paren{\xi}$ leads 
\begin{align}
  n_a \paren{\xi}
 &= 2 |a|  f^{-2}_{a} \tilde{\rho}^{t}_{a} \paren{0,\xi}
\nonumber\\
 &= 2 |a|   A_{1} \paren{0,\xi} 
  f^{-1}_{a} \bigl( f_{a+1}^{-1} -f_{a-1}^{-1} \bigr) , 
\quad (a < a_* ).
\end{align}
The total particle density \eqref{eq:n2} is 
\begin{align}
  n\paren{\xi} 
  = 1 - ~\tilde{\rho}^{t}_{-\infty} \paren{0,\xi}
  = 1 - 2 \sinh |\bar{\mu} | A_{1} \paren{0,\xi}.
\label{eq:n_xi}
\end{align}
Here, we have used Eq.~\eqref{eq4:rhotp} in the limit $a\to -\infty$.
Since $n (\xi ) = \sum_{a \le 0} n_a (\xi)$, 
$A_{1}(0,\xi)$ is given as 
\begin{align}
    A_{1}\paren{0,\xi}
    = C \paren{a_{*},| \bar{\mu}|} 
    \left[ 1 - \sum_{a_{*} \le a \le 0} 
    n_{a} \paren{\xi} \right],
\end{align}
with
\begin{align}
 C \paren{a_{*},| \bar{\mu}|}^{-1} 
& = 
   2 \sinh |\bar{\mu}| - 
   \sum_{a < a_{*}} 2a f^{-1}_{a} 
   \bigl( f^{-1}_{a+1} - f^{-1}_{a-1} \bigr) 
\nonumber\\
&=2f^{-1}_{a_{*}}\bce{-a_{*}f^{-1}_{a_{*}-1}+\cosh\left[(a_{*}-1)|\bar{\mu}|\right]},
\label{eq:funcc}
\end{align} 
where we have used the identity
$\sum_{a\leq a_{*}}f^{-1}_{a} f^{-1}_{a-1}=e^{(a_{*}-1)| \bar{\mu}|}f^{-1}_{a_{*}}$.
This result shows that $A_{1} (0, \xi )$ is finite. 

The results above are obtained for the region
$V_{\LL} < \xi < \xi_{-\infty}^{-}$.  
When $\bar{\mu}=0$, the initial left equilibrium has the 
particle density $n^{\LL }=1$. 
Equation~\eqref{eq:n_xi} shows 
that the particle density remains unity in this $\xi$-region 
\begin{align}
    n(\xi)=1 , 
   \quad  
   \bigl( V_{\LL} < {}^\forall \xi < \xi_{-\infty}^{-} \bigr).  
\end{align}
This identity does not depend on the value of $A_{1} (0, \xi )$, and thus the result holds for 
any initial state in the right part.
The continuity equation implies $\frac{\diff}{\diff \xi}j_{n}=0$, 
and since the boundary value vanishes $j_n (V_{\LL})=0$, 
the particle density current also vanishes within this whole 
region 
\begin{equation}
  j_n (\xi ) = 0 , 
   \quad  
   \bigl( V_{\LL} < {}^\forall \xi < \xi_{-\infty}^{-} \bigr).  
\end{equation}

Note that from Eq.~\eqref{eq:dvine} and Eq.~\eqref{eq:leftslope}, we have proved the existence of a clogged region. The width of this region is $\xi^{-}_{-\infty}-V_{\LL}$, and the value of $\xi_{-\infty}^{-}$ depends on the initial state in the right part. To determine $\xi^{-}_{-\infty}$, one needs $\rho^{t}_{-\infty}\paren{\clam,\xi}$ and should go back to solve the whole set of Eqs.~\eqref{eqD:te1}-\eqref{eqD:te4}, which includes $\vartheta^{\RR}_{a}$ in the initial right state, and it is difficult to solve them analytically.

We can repeat similar calculations for the magnetization.
Let us now consider the case $B_{\RR}\neq 0$ and the $\xi$ region 
\begin{align}
    V_{\LL} < \xi<\xi_{\infty}^{-},
\label{eq:xi2}
\end{align}
and any $\xi$ in this region has a positive
integer $a_{\star}$ such that $\xi < \xi_{a}^{-}$, 
${}^\forall a > a_{\star} $. Whether $\xi^{-}_{-\infty}$ or $\xi^{-}_{\infty}$ is larger depends on the initial conditions.
The quasiparticles' contributions to magnetization are 
similarly calculated as 
\begin{equation}
  m_{a} (\xi ) 
  = 2a f^{-1}_{a} \bigl( f^{-1}_{a+1} - f^{-1}_{a-1} \bigr) 
  C \bigl( a_{\star},\bar{B} \bigr) 
  \sum_{a=0}^{a_{\star}}
   m_{a} (\xi ) ,
\end{equation}
for $ a>a_{\star}$ 
with 
\begin{align}
m_{0}(\xi)&\equiv \tilde{n}_{0}(\xi),\nonumber\\
m_{a}(\xi)&\equiv -a\tilde{n}_{a}(\xi),\quad (a>0),
\end{align}
and 
\begin{equation}
C (a_{\star}, \bar{B})^{-1}
=2f^{-1}_{a_{\star}}\bce{a_{\star}f^{-1}_{a_{\star}+1}+\cosh\left[(a_{\star}+1)\bar{B}\right]}. 
\end{equation} 
The magnetization $m(\xi)=\sum_{a\geq 0}m_{a}(\xi)$ is given as
\begin{equation}
   m (\xi )
  = 2  \sinh \bar{B} ~C ( a_{*},\bar{B} )
   \sum_{a=0}^{a_{*}} m_{a} \paren{\xi}.
\label{eq:m_xi}
\end{equation}
In the case of $\bar{B}=0$, i.e. the initial magnetization 
vanishes in the left part $m^{\LL }=0$, the magnetization remains zero 
in the spin clogged region 
\begin{align}
    m(\xi )=0 , 
    \quad
    \bigl( V_{\LL} < {}^{\forall} \xi < \xi_{\infty}^{-} \bigr),  
\end{align}
and no spin current flows 
\begin{align} 
    j_m (\xi ) = 0 , 
    \quad 
    \bigl( V_{\LL} < {}^{\forall} \xi < \xi_{\infty}^{-} \bigr).
\end{align}
A similar phenomenon is known in the XXZ model~\cite{PhysRevB.96.115124}.

For $\bar{\mu} \ne 0$ and $\bar{B} \ne 0$, 
there is no clogged region for both $n(\xi)$ and $m (\xi)$, 
and $j_n$ and $j_m$ both start to flow at $\xi = V_{\LL}$. 
We can show their general relation 
in the more restricted $\xi$-region 
\begin{equation}
 V_{\LL} < \xi < V_{\LL ,1} \equiv \min_{a \ne 0} \xi_a^{-}. 
\end{equation}
In this region, 
Eq.~\eqref{eq:n_xi} and Eq.~\eqref{eq:m_xi} are both satisfied 
with $a_{*}=a_{\star}=0$, 
and therefore 
\begin{align}
    n(\xi)&=1-\tanh |\bar{\mu}| \bigl[ 1-n_{0}\paren{\xi} \bigr],\\
    m(\xi)&=\tanh \bar{B}\cdot m_{0}\paren{\xi}.
\end{align}
Eliminating $n_0 (\xi)$ by using the relation 
$n_{0}=2m_{0}$, one derives a relation between 
$n\paren{\xi}$ and $m\paren{\xi}$ 
\begin{align}
   \frac{ 1-n\paren{\xi} }{ \tanh |\bar{\mu}| } 
 + \frac{ 2 m\paren{\xi} }{ \tanh \bar{B} }=1, 
 \quad 
  \bigl( V_{\LL} < {}^{\forall} \xi < V_{\LL ,1} \bigr) . 
\end{align}
This also leads to the relation between the corresponding currents
\begin{align}
  \frac{  j_{n}\paren{\xi} }{ \tanh |\bar{\mu}| }  
 =\frac{ 2j_{m}\paren{\xi} }{ \tanh \bar{B} }, 
 \quad 
  \bigl( V_{\LL} < {}^{\forall} \xi < V_{\LL ,1} \bigr) . 
\label{eq:jnandjm}
\end{align}
Here, we have used the boundary values 
$j_n (V_{\LL}) =0$ and $j_m (V_{\LL}) =0$. 

Finally, we show that the energy current is nonzero in the charge clogged region $V_{\LL}<\xi<\xi^{-}_{-\infty}$, where the particle density current vanishes. Despite the constant particle density, the contribution of each type of quasiparticles changes in this $\xi$-region. Since their bare energy depends on quasiparticle type, it is expected that the energy current flows while the particle density current vanishes. We can explicitly show a nonvanishing $j_{e}$ in the restricted $\xi$-region $V_{\LL}<\xi<V_{\LL,1}$ in the special case of $|\bar{\mu}|=\bar{B}$, where the chemical potential and the magnetic field are canceled for spin-up electrons: $\bar{\mu}+s_{\uparrow}\bar{B}=0$. This is shown in Appendix~\ref{app:iene}.

\section{Numerical GHD solution at finite temperatures}\label{sec:5}
Now, we present numerical solutions for 
the generalized hydrodynamics and investigate 
the density and current profiles.  
In this section, we use a simple case of the partitioning protocol 
that the right part has no electron at $t=0$.  
Thus, the corresponding quasiparticle fillings 
are zero $\vartheta_a^{\RR} = 0$ 
in Eq.~\eqref{eq:vartheta}.
For the initial left state characterized by 
$\beta_{\LL}$, $\bar{\mu} = \beta_{\LL} \mu_{\LL}$, 
and $\bar{B} = \beta_{\LL} B_{\LL}$,
we need to obtain the filling functions $\bce{\vartheta^{\LL}_{a}}$ 
and they are determined by
solving the TBA equations \eqref{eq:tba4}-\eqref{eq:tba3}
and Takahashi equations \eqref{eqD:te1}-\eqref{eqD:te4}.   
Since they both consist of infinitely many coupled equations, 
we need a cut-off $a_c$ for the string length 
in their numerical calculations. Note that for at high temperatures with small $|\mu_{\LL}|$, 
the initial left particle density of $k$-$\clam$ string $n^{\LL}_{a<0}$ becomes large, therefore, we need a large $a_{c}$ in our calculations.
Simple setting 
$\eta_a (\clam )= 0$ and $\rho_a^t (\clam ) = 0 $ 
for $|a|>a_c$ is not consistent with the boundary conditions of the TBA equations Eq.~\eqref{eq:tba5} and 
the expression of $n$ Eq.~\eqref{eq:n2},
and we use better approximations explained below.  

As for the TBA equations, we have employed the approximation 
used in Ref.~68\nocite{PhysRevB.65.165104}. 
For large strings with $|a|>a_{c}$, 
we approximate $s\paren{\clam}$ by $\frac{1}{2}\delta\paren{\clam}$ 
in Eqs.~\eqref{eq:tba4} and obtain 
\begin{align}
   \eta_{a}\paren{\clam}
  \approx [1+\eta_{a-1}\paren{\clam}]
    [1+\eta_{a+1}\paren{\clam}],
    \quad 
    (|a|>a_{c}).
\label{eq:aptba}
\end{align}
Considering the asymptotic behavior \eqref{eq:tba5}, its solution 
is given as  
\begin{equation}
 \eta_{a}\paren{\clam} 
=\left\{
\begin{array}{ll}
\displaystyle 
\frac{\sinh^2 [ ( g_{-} \paren{\clam} +|a| ) \bar{\mu} ]}
     {\sinh^2 \bar{\mu} }
-1,
& (a < - a_c),
\\[10pt]
\displaystyle 
\frac{\sinh^2 [ ( g_{+} \paren{\clam} + a  ) \bar{B} ]}
     {\sinh^2 \bar{B} }
-1,
& (a >  a_c).
\end{array}
\right. 
\label{eq:etasol2}
\end{equation}
In particular, when $\bar{\mu}=0$ or $\bar{B}=0$, 
\begin{equation}
 \eta_{a}\paren{\clam} 
=\left\{
\begin{array}{ll}
\displaystyle 
\paren{ g_{-} \paren{\clam} +|a| }^{2}
-1,
& (\bar{\mu}=0,~a < - a_c),
\\[10pt]
\displaystyle 
\paren{ g_{+} \paren{\clam} + a  }^{2}
-1,
& (\bar{B}=0,~a >  a_c).
\end{array}
\right. 
\end{equation}
Here, two functions $g_{\mp}$ are determined by the boundary values 
at $a = \mp a_c$
\begin{align}
   g_{\mp} \paren{\clam}
   &= 
   \frac{1}{\gamma}  \, 
  \sinh^{-1} 
  \left[\sqrt{1+\eta_{ \mp a_{c}} \paren{\clam}} \sinh \gamma  \right]
  - a_{c},
\end{align}
with $\gamma= \bar{\mu}$ for $g_{-}$ and $\bar{B}$ for $g_{+}$.  
Thus, the TBA equations are now closed for 
$(2a_{c}+1)$ unknown functions 
$\{ \eta_a (\clam ) \}$ $(-a_c \le a \le a_c)$, 
and we numerically solve them by iteration. 

As for Takahashi's equations, we have used another 
simple approximation 
$ \rho^{t}_{a} \paren{\clam} = \rho^{t}_{-\infty} \paren{\clam}$ 
$(a \le - a_c + 1)$ and 
$ \rho^{t}_{a} \paren{\clam} = \rho^{t}_{\infty} \paren{\clam}$ 
$(a \ge  a_c - 1)$.  
This is based on the fact that $\rho_a^t (\clam )$ converges 
smoothly to $\rho_{\pm \infty}^t (\clam )$ as $a \rightarrow \pm \infty$. 
Actually, if a string $a$ is so long such that 
$\vartheta_a (\clam ) \approx \vartheta_{a \pm 1} (\clam ) \approx 0$, 
its distribution follows the recurrence relation 
$
   \rho^{t}_{a} \paren{\clam} 
  \approx 
   \bigl[ \rho^{t}_{a-1} \paren{\clam} + \rho^{t}_{a+1} \paren{\clam} 
   \bigr] /2 
$
and this is consistent with this approximation, 
and we assume that the difference from $\rho_{\pm \infty}^t (\clam )$ 
remains small at the cut-off $\pm a_c$.  
Thus, Takahashi's equations are now reduced to be about 
$( 2a_{c} - 1 )$ unknown functions 
$\{ \rho^{t}_{a} \paren{w} \}$ $( - a_c +1 \le a \le a_c - 1)$. 
We numerically solve the linear integral equations with limiting the functions' domain. 
As for the derivatives of dressed energies $\bce{\dre{a}\paren{w}}$,
we approximate the equations~\eqref{eq:de4}-\eqref{eq:de3} in a similar way and numerically solve them. 
\begin{figure}[tb]
\centering
\includegraphics[width=\columnwidth,clip]{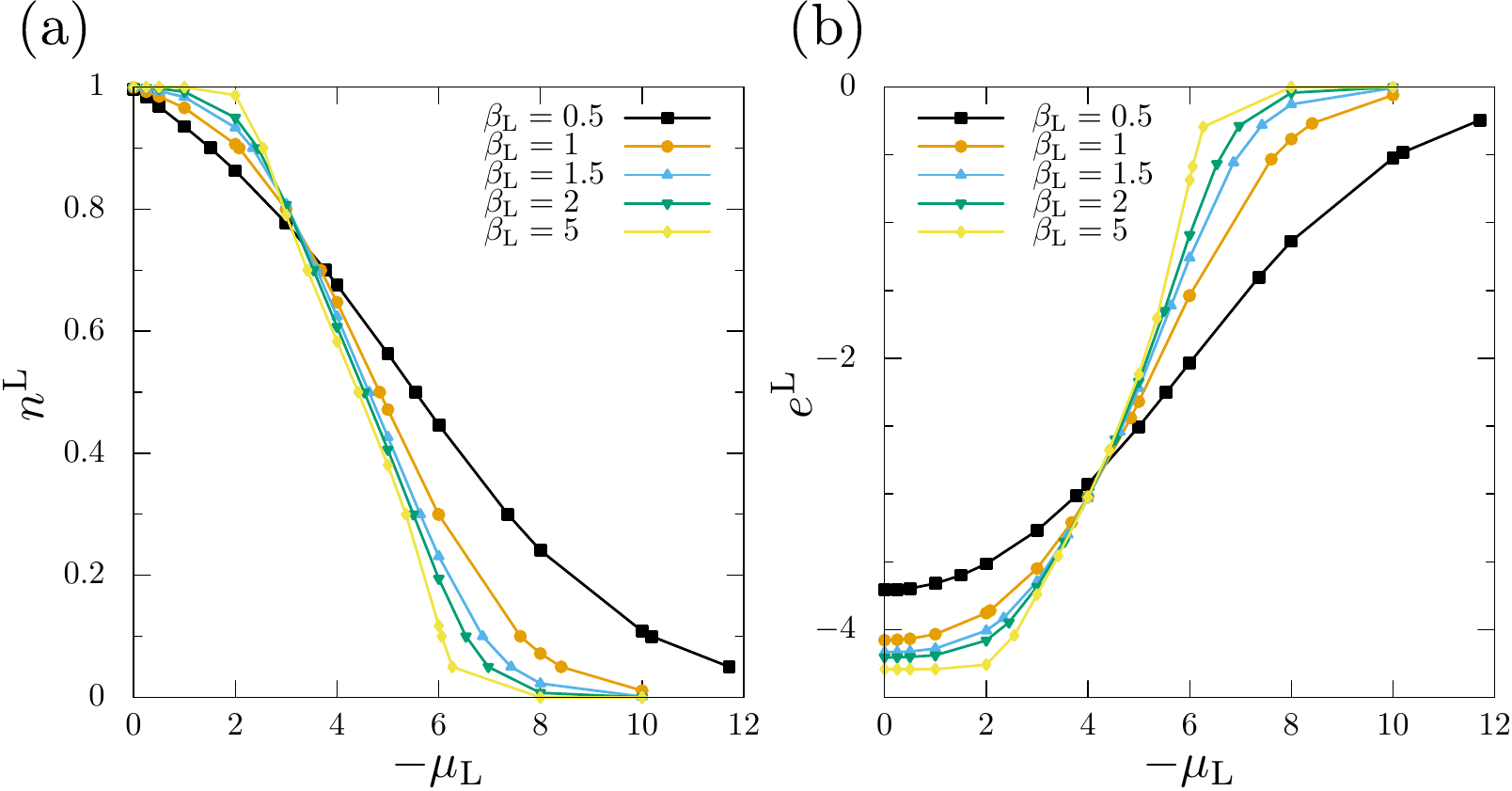}
\caption{
Chemical potential dependence 
of (a) particle density 
and (b) energy density in the initial left equilibrium state.
Five curves correspond to different temperatures 
$0.5 \le \beta_{\LL} \le 5$.  
}
\label{fig:th}
\end{figure}
\begin{figure}[t]
\centering
\includegraphics[width=\columnwidth,clip]{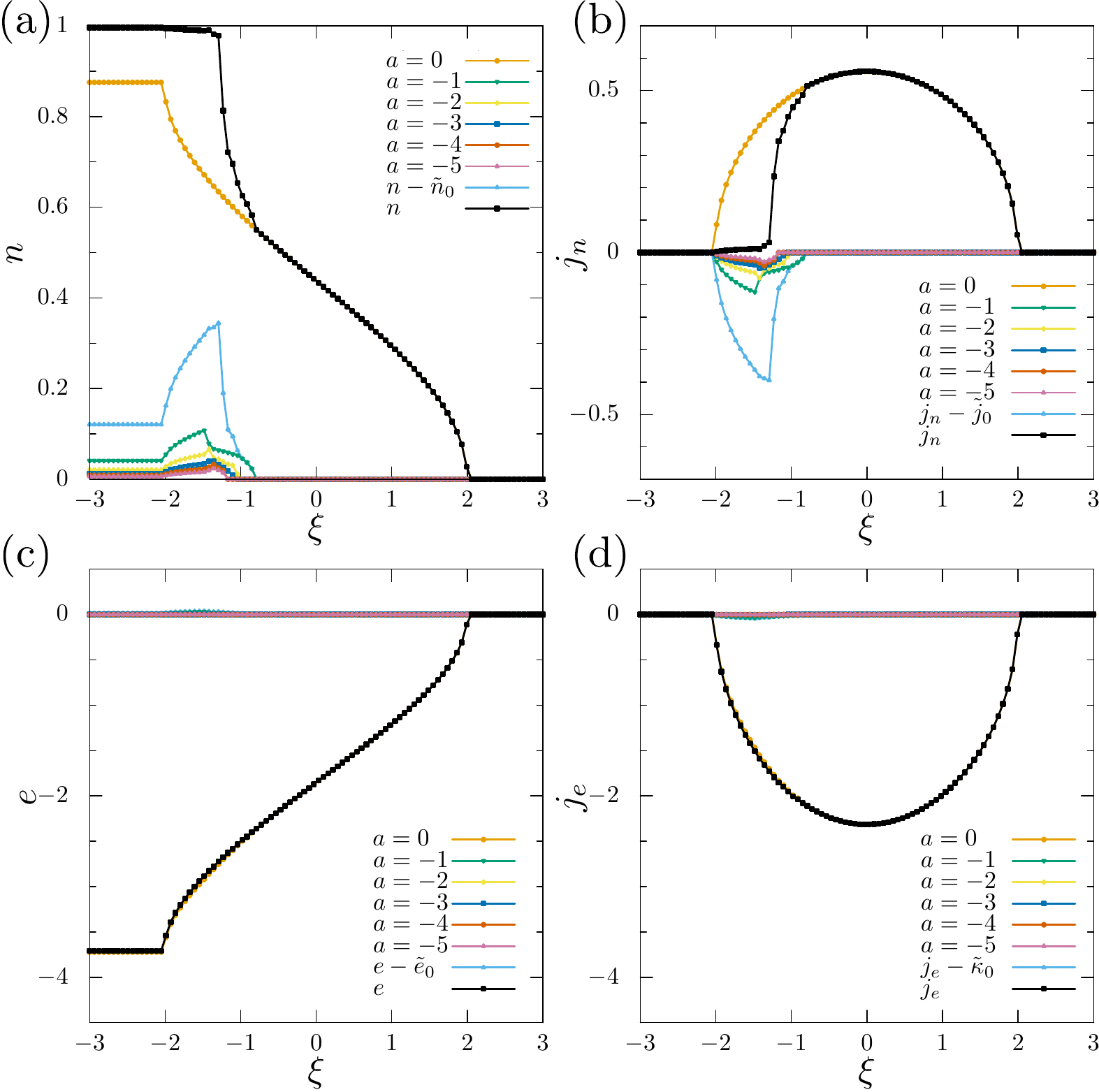}
\caption{
Profiles of particle and energy densities and their currents: 
(a) $n(\xi )$, (b) $j_n (\xi )$, (c) $e (\xi )$, and (d) $j_e (\xi)$, 
as functions of the ray $\xi=x/t$. 
The initial condition is $(\beta_{\LL},\mu_{\LL})=(0.5,0)$. 
Black lines show the total values of each quantity with the cut-off $a_c=48$. 
Orange lines are contributions of scattering states ($a=0$). 
Light blue lines are the sum of contributions of bound states
($ -47 \le a \le -1 )$.  
The other lines are contributions of quasiparticles 
with $-5 \le a \le -1$.
}
\label{fig:st}
\end{figure}

Details of numerical calculations are as follows.
For solving both filling functions and dressed quantities, convolutions include $s(\clam )$ in their kernel.
This function decays exponentially at large $|\clam |$ and
we use the cutoff for integration region
$-20 u \le \clam \le 20 u$. Domain of functions to be solved is limited to $-200u \leq \clam \leq 200u$. 
For solving dressed quantities,
we calculate integrals including filling functions
$\bce{\vartheta_{a}(w,\xi)}$.
Each $\vartheta_{a}$ is not continuous
but jumps at the points on the curve $\xi=\ddd{v}{a} ( w ,  \xi )$.
Therefore, we first determine these discontinuous points
and minimize numerical errors in integration
by taking account of the jump in the integrand.
In each iteration, the dressed velocities are updated from
their previous values.
Therefore, we need to determine discontinuous
points each time, but this accelerates convergence
and improves accuracy.
The number of iterations for solving dressed quantities
is set at most 100 times.
When the results did not converge within this limit,
we estimated error bars
for local conserved quantities and currents
from their values in the last 20 iterations.

\begin{figure}[b]
\centering
\includegraphics[width=\columnwidth,clip]{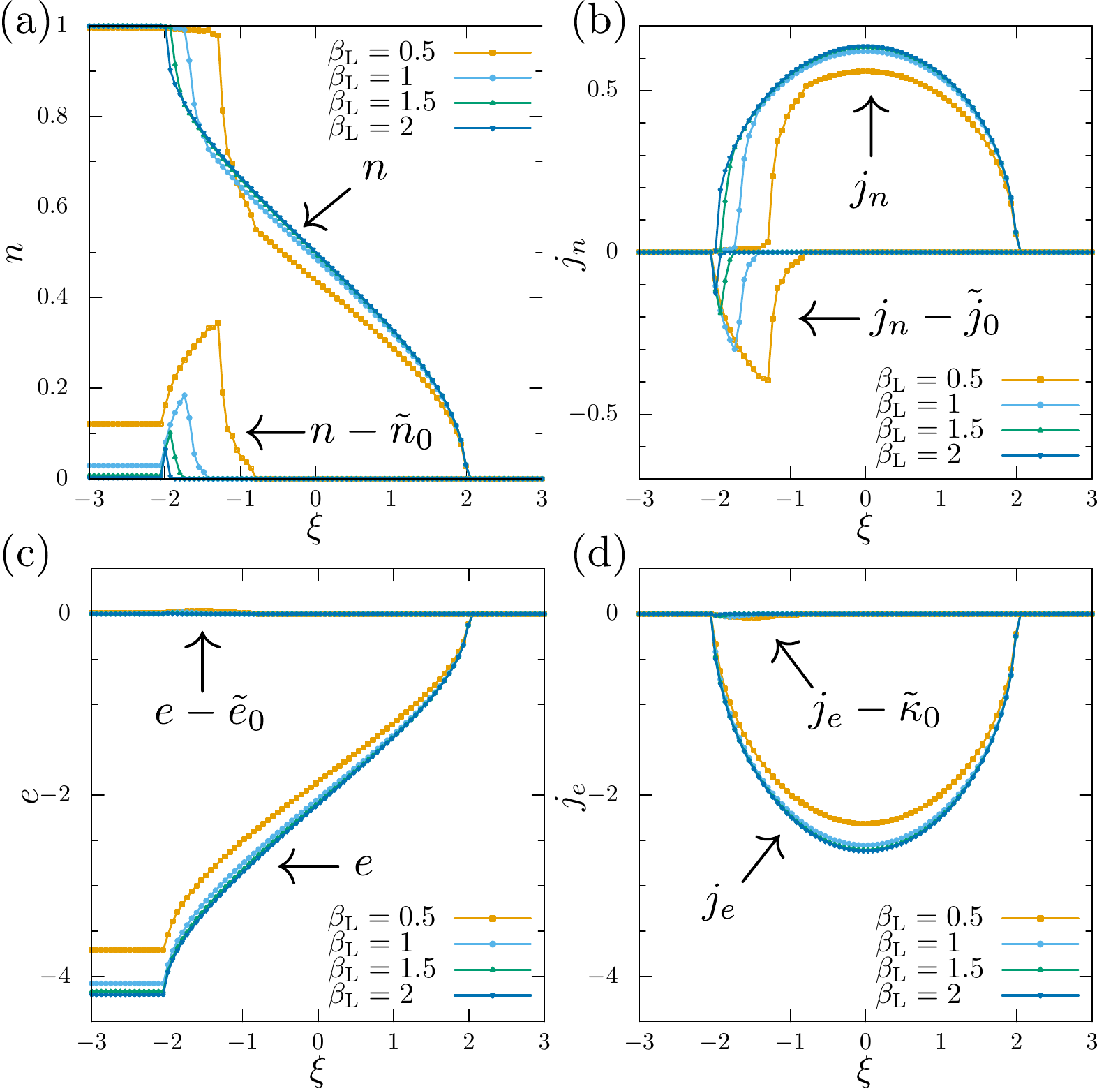}
\caption{
Effects of varying the temperature $0.5 \le \beta_{\LL} \le 2$ 
on quasiparticle contributions in 
(a) $n(\xi )$, (b) $j_n (\xi )$, (c) $e (\xi)$, and (d) $j_e (\xi )$.  
The initial chemical potential is fixed at $\mu_{\LL} = 0$.  
For each temperature, the total value of density or current 
and the sum of bound state contributions are separately plotted.  
}
\label{fig:betasumqp}
\end{figure}

\subsection{Particle and energy  densities and their currents}
In our numerical calculation, we set the repulsion $u=2$ and
the cut-off $a_{c}=48$.
We analyzed particle density $n(\xi)$, energy density $e(\xi)$, and their currents $j_n (\xi )$ and $j_{e}(\xi)$ at zero magnetic field $B_{\LL}=0$.
First, to clarify the correspondence of initial states to chemical potential and temperature, Fig.~\ref{fig:th} (a) and (b) show the $\mu_{\LL}$ dependence of particle density and energy density in the initial left equilibrium state~\cite{PhysRevB.65.165104}.

The profiles of $n\paren{\xi}$ and $j_n (\xi )$ are shown in Fig.~\ref{fig:st} (a) and (b).
Contributions of different types of quasiparticles to particle density $n_{a}\paren{\xi}$
and their currents $j_{n,a} (\xi )$ are also plotted in the same panels.  
The particle density and its current show a similar 
behavior as those in the infinite-temperature 
limit discussed in Sec.~\ref{sec:4}.  
For initial states at or near half filling $\mu_{\LL} \approx 0$, 
there appears a clogged $\xi$-region where 
the particle density hardly changes from the left bulk part 
\begin{equation}
n (\xi ) \approx 1 ,   
\quad ( \bar{\mu} \approx 0, \ V_{\LL} < \xi < \xi^{-}_{-\infty} ) . 
\end{equation}
Analysis of different types of quasiparticles 
reveals that scattering states (i.e. real-$k$ type) have 
a dominant contribution $n_0 (\xi) \gg n_{a<0} (\xi )$
and that it starts to decrease noticeably already at 
$\xi = V_{\LL}$. 
However, this decrease is nearly compensated by the increase 
in the contributions of bound states (i.e., $k$-$\Lambda$ strings)  
$\{ n_a (\xi )\}_{a<0}$, in particular those of $-5 \le a \le -1$.  
For $\xi > \xi_{-\infty}^{-}$, these bound state contributions 
are vanishingly small, and the total particle density $n (\xi )$ 
shows a large decrease with increasing $\xi$.  
In other words, the charge current is mainly carried 
by scattering states 
but it is nearly canceled by counterflow carried by bound states 
in the clogged $\xi$-region.
\begin{figure}[tb]
\centering
\includegraphics[width=5.0cm,clip]{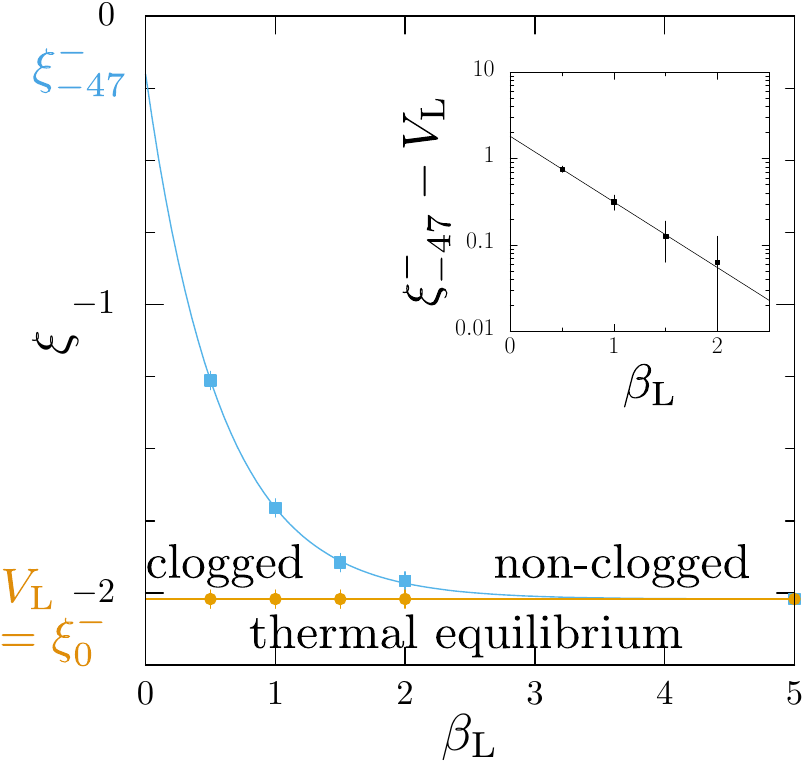}
\caption{
Clogged region at finite temperatures $0.5\leq \beta_{\LL} \leq 5$.
The cut-off is $a_{c}=48$, and $\xi^{-}_{-47}$ is an approximate value of $\xi^{-}_{-\infty}$.
The initial chemical potential is fixed at $\mu_{\LL}=0$. The blue line is a fitting function $\xi^{-}_{-47}-\xi^{-}_{0}=a\ b^{-\beta_{\LL}}$, where $a=1.82$ and $b=5.76$. Inset: semi-log plot of $\beta_{\LL}$ dependence of the width.
}
\label{fig:ray}
\end{figure}
\begin{figure}[tb]
\centering
\includegraphics[width=7.0cm,clip]{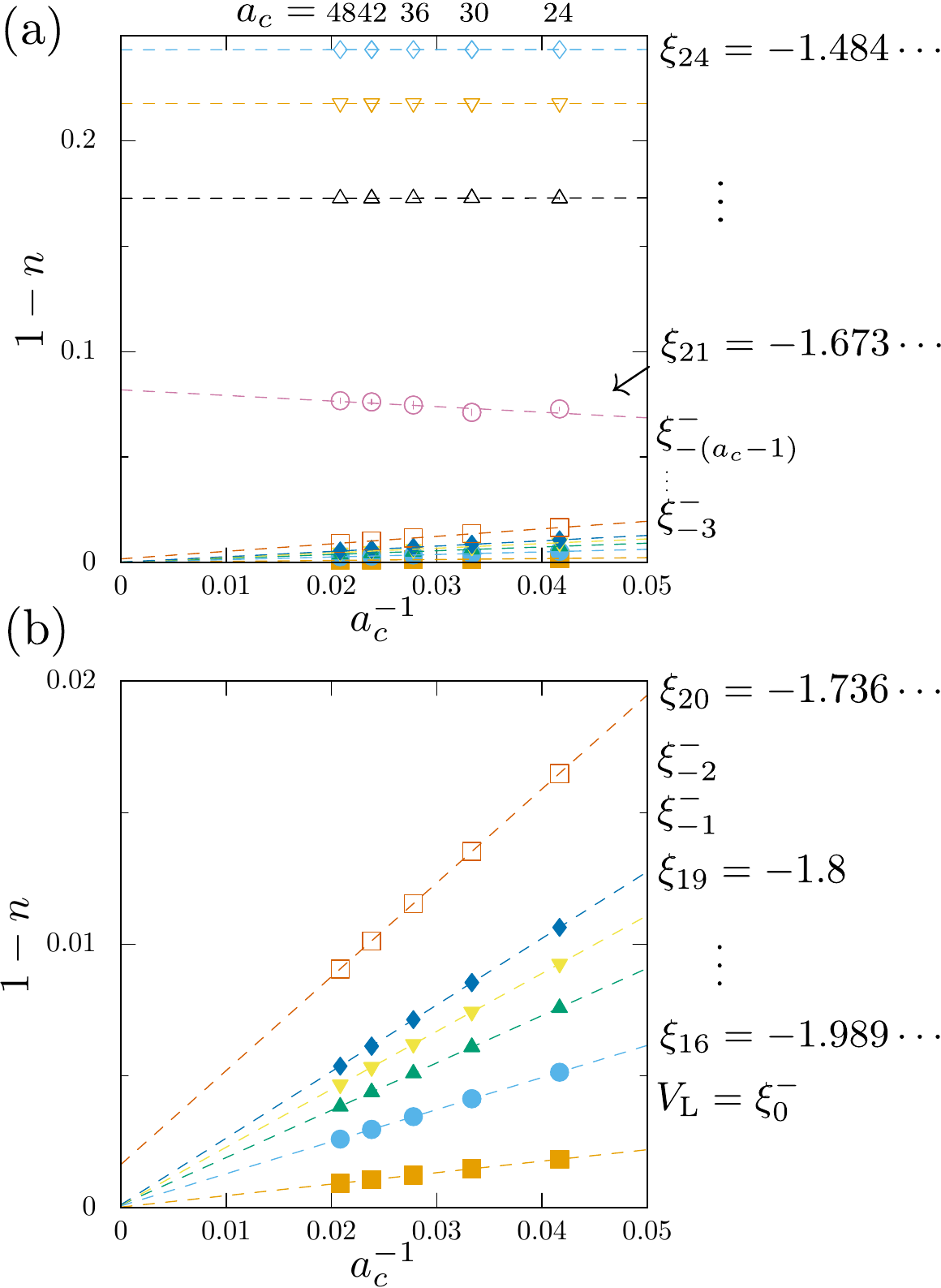}
\caption{
Extrapolation for checking if $n(\xi ) =1$ 
in the limit of the cut-off $a_c \rightarrow \infty$.  
The initial condition is $(\beta_{\LL},\mu_{\LL})=(1,0)$. 
(a) 
Analysis for those $\xi$ points in or near the clogged region:
$\xi_{l}=-3+6l/95~(15\leq l \leq 24)$.
Symbols show the numerical data, and their linear fittings 
are plotted by dashed lines.  
(b) 
Magnified plot of the lower part in the panel (a).  
Light cones $\xi^{-}_{a}$ in this $\xi$ range are 
also marked roughly at the place corresponding to their values.  
Notice $\xi_{0} < \cdots < \xi_{24} <\xi^{-}_{a}$ for all $a>0$.
} 
\label{fig:ac}
\end{figure}
\begin{figure}[tb]
\centering
\includegraphics[width=7.0cm,clip]{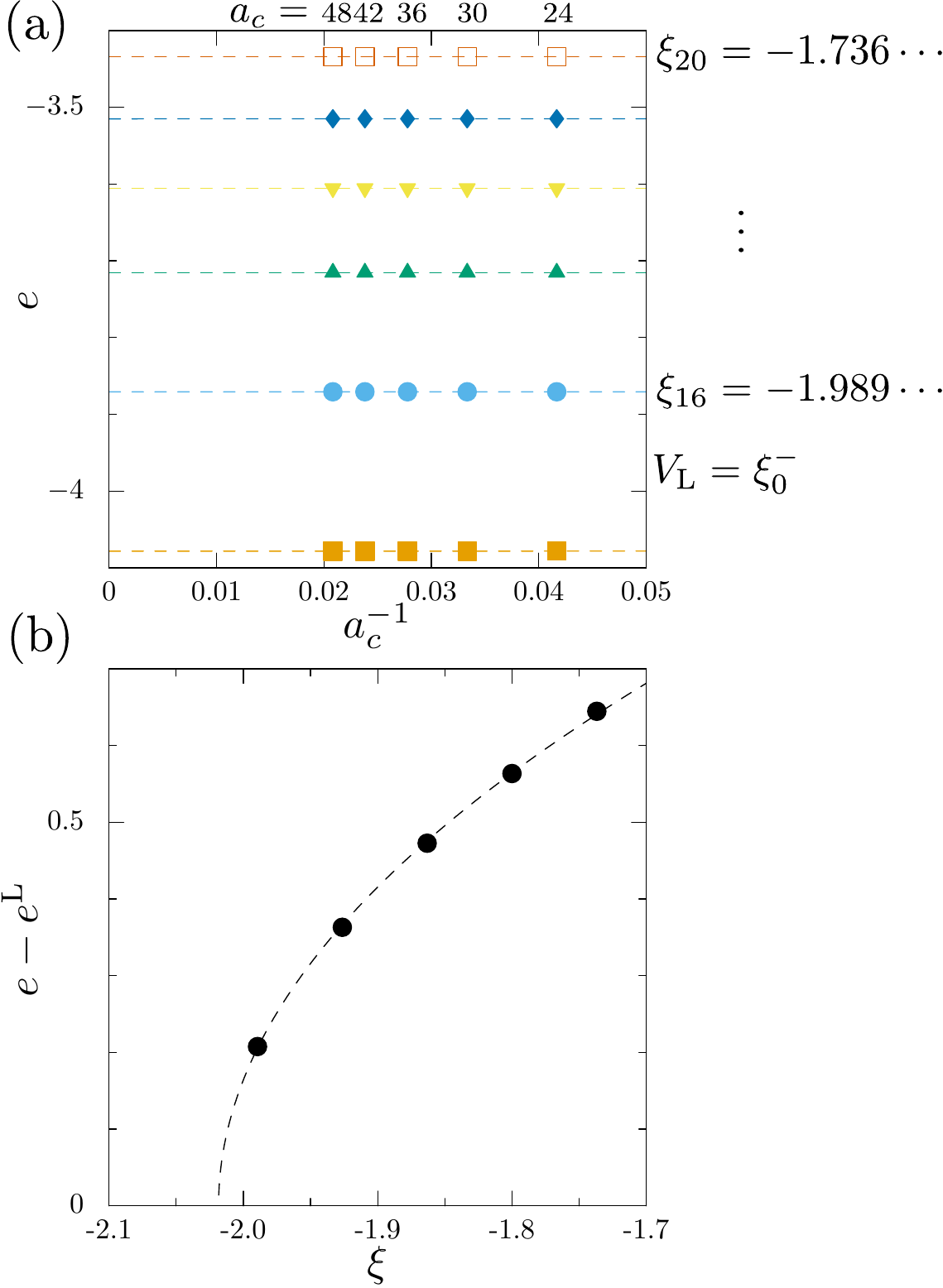}
\caption{
Convergence check of energy density in the limit of 
the cut-off $a_c \rightarrow \infty$.  
(a) 
Analysis for those $\xi$ points in or near the clogged region:
$\xi_{l}=-3+6l/95~(15\leq l \leq 20)$.
Symbols show the numerical results, and their linear fittings 
are plotted by dashed lines.  
The initial condition is $(\beta_{\LL},\mu_{\LL})=(1,0)$. 
(b) 
Singularity of energy density. 
Symbols are extrapolated values determined in the panel (a).
The dashed line is a fitting function $a\sqrt{\xi-b}$,
where $a=1.21$ and $b=-2.02$.
}
\label{fig:eac}
\end{figure}

Figure~\ref{fig:betasumqp} shows contributions of quasiparticles at several temperatures with $\mu_{\LL}=0$ fixed. For each temperature, one line shows the total density or current and the other line shows the sum of contributions of bound states. Even at lower temperatures, the clogged region exists, but its width shrinks. 
Figure~\ref{fig:ray} shows the $\beta_{\LL}$ dependence of the width. $\xi^{-}_{a}$ is determined by comparing $\xi$ and $\min_{w}\ddd{v}{a} ( w ,  \xi )$ at each $\xi$. We checked that the cut-off $a_{c}=48$ is large enough to calculate $\xi^{-}_{-\infty}$. In addition, the $\beta_{\LL}$ dependence of $V_{\LL}$ is negligible. The inset show that the width decreases exponentially with increasing $\beta_{\LL}$.

\begin{figure}[tb]
\centering
\includegraphics[width=\columnwidth,clip]{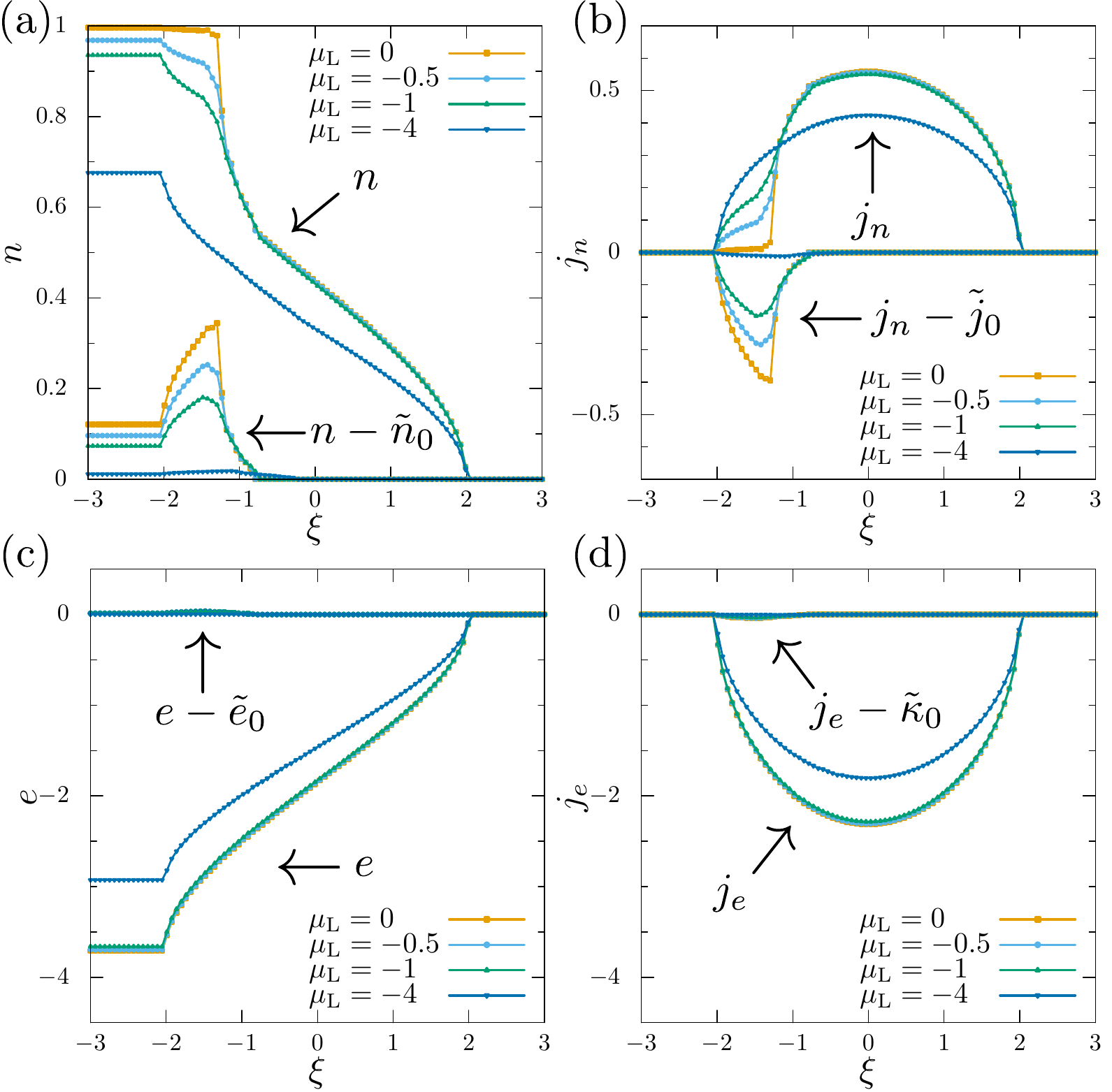}
\caption{
Effects of varying the chemical potential $-4 \le \mu_L \le 0$ on 
quasiparticle contributions in 
(a) $n(\xi)$, (b) $j_n (\xi )$, (c) $e (\xi)$, and (d) $j_e (\xi )$.  
The initial temperature is fixed at $\beta_{\LL}=1$. 
For each $\mu_{\LL}$, the total value of density or current 
and the sum of bound state contributions are separately plotted.  
}
\label{fig:musumqp}
\end{figure}

To study in detail the deviation of particle density from unity in the clogged region due to temperature change, we analyzed its convergence with increasing cut-off $a_{c}$. Figure~\ref{fig:ac} shows $1-n$ 
as a function of $a_{c}^{-1}$
at $\mu_{\LL}=0$ and $\beta_{\LL}=1$ for several $\xi$'s around the clogged region together with linear fitting. The extrapolation to $a_{c}\to \infty$ shows that the particle density is pinned as $n (\xi ) = 1$ in the clogged region even at finite temperatures. Figure~\ref{fig:eac}~(a) shows the energy density $e(\xi)$ plotted in the same way. It shows the energy density has already converged
with this cut-off $a_{c}$. The energy density $e(\xi)$ has a singularity 
\begin{align}
e\paren{\xi}-e^{L}\sim \sqrt{\xi-V_{\LL}}, \quad (\xi\geq V_{\LL}),
\label{eq:esin}
\end{align}
close to $V_{\LL}$
as shown in Fig.~\ref{fig:eac}~(b). Here, $e^{\LL}$ is the initial left energy density. 
The continuity equation \eqref{eq:cont} leads to the singularity of energy current 
\begin{align}
j_{e}\paren{\xi}\sim 3V_{\LL}\sqrt{\xi-V_{\LL}}+\paren{\xi-V_{\LL}}^{3/2}.
\end{align}
Figure~\ref{fig:musumqp} shows the $\mu_{\LL}$ dependence of contributions of quasiparticles with $\beta_{\LL}=0.5$ fixed. In $V_{\LL} < \xi < \xi^{-}_{-\infty}$, the small differences of $\mu_{\LL}$ from $0$ change the contributions of the bound states to particle density more than those of the scattering states. Consequently, the clogged region vanishes, and the particle density has a singularity $n^{\LL}-n(\xi)\sim \sqrt{\xi-V_{\LL}}$ close to $V_{\LL}$, where $n^{\LL}$ is the initial left particle density.
As for the particle density for $\mu_{\LL}=0$, $\bce{n_{a}(\xi)}$ show a singularity similar to Eq.~\eqref{eq:esin}, $n_{a}(\xi)-n^{\LL}_{a}\sim \pm\sqrt{\xi-V_{\LL}}$ for $a\leq 0$, where $n_{a}^{\LL}$ is each initial left particle density, although, their contributions are canceled out.

Energy density and energy current behave differently from $n\paren{\xi}$ and $j_{n}\paren{\xi}$, 
and $j_{e}\paren{\xi}$ shows no suppression in the region 
$V_{\LL} < \xi < \xi_{-\infty}^{-}$.  
All the contributions of different quasiparticle types,  
$ \tilde{e}_{0} $ and $\{ \tilde{e}_{a<0} \}$,  
start to increase at the same position $\xi = V_{\LL}$, 
and their currents 
$ \tilde{\kappa}_{0} $ and $\{ \tilde{\kappa}_{a<0} \}$ 
flow in the same direction. 
As in the particle density current, 
the contribution of scattering (i.e., real-$k$) states 
is dominant, 
$ | \tilde{\kappa}_{0} | \gg | \tilde{\kappa}_{a} | $ $({}^\forall a < 0 )$.   
Consequently, in the clogged region 
$V_{\LL} < \xi < \xi_{-\infty}^{-}$,
particle density current is suppressed nearly completely,  
while energy current is large, 
when the initial density is near half filling 
$\beta_{\LL} \mu_{\LL} \approx 0$. 
This asymmetry between charge and energy currents
is quite different from the one expected from 
Wiedemann-Franz law in thermal equilibrium~\cite{test}.  

\begin{figure}[tb]
\centering
\includegraphics[width=\columnwidth,clip]{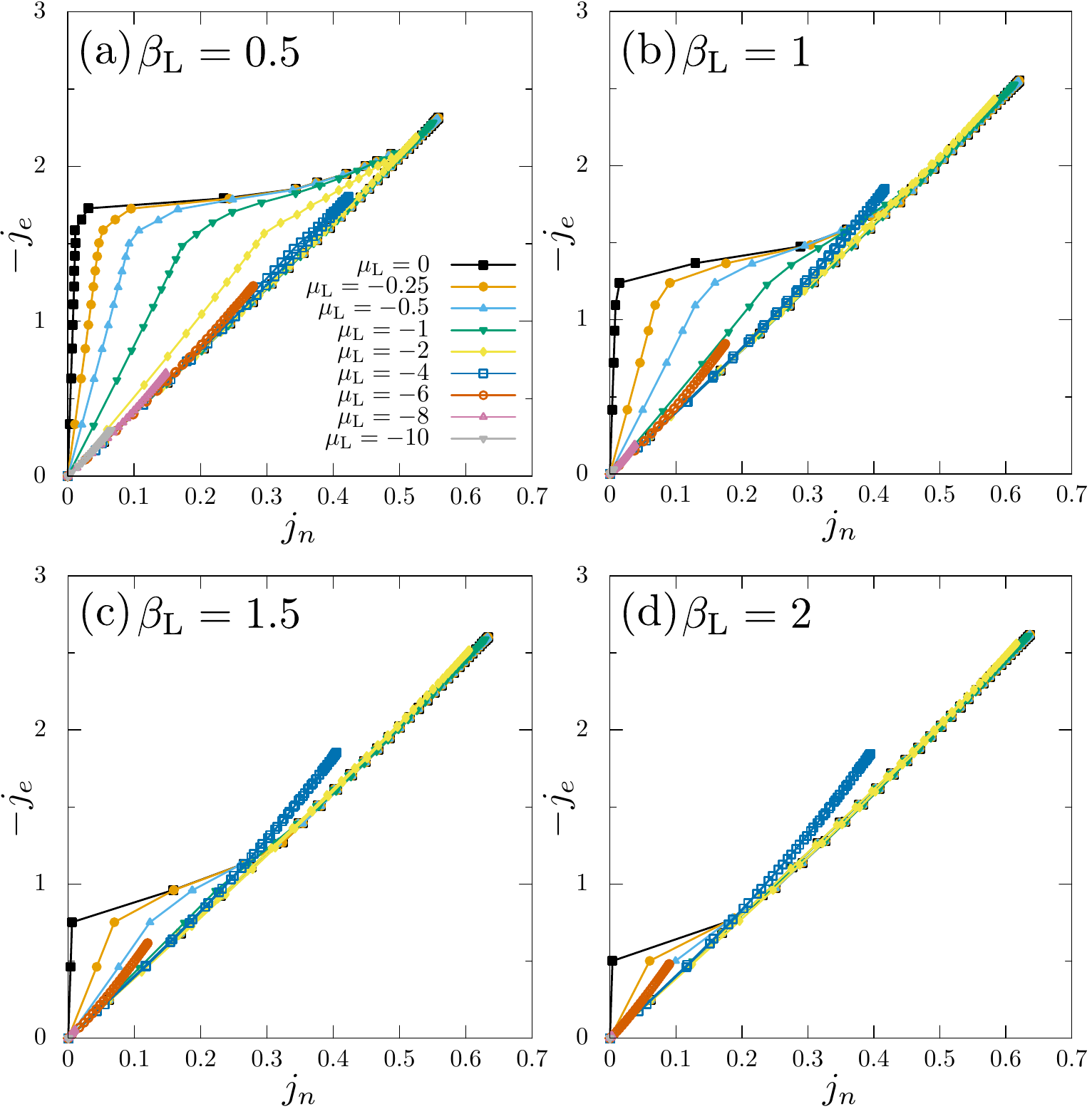}
\caption{
Relation of particle density current $j_n (\xi )$ and energy current 
$j_e (\xi )$ plotted for $\xi=\xi_{l}=-3+6l/95, (l=0,1,\cdots,95)$. 
The four panels correspond to different values 
of the initial temperature: 
$\beta_{\LL}=$ (a)~$0.5$, (b)~$1$, (c)~$1.5$, and (d)~$2$.  
Nine data sets for $-10 \leq \mu_{\LL} \leq 0$ 
are plotted in each panel.  
The points $|\xi_{l}| \le 2$ are within some light cones, 
and the corresponding filling functions differ 
from those in the initial states.
}
\label{fig:jnje}
\end{figure}

Figure~\ref{fig:jnje} compares the energy currents and 
the particle density currents in the region of $-3 \leq \xi \leq 3$ 
for various values of $\beta_{\LL}$ and $\mu_{\LL}$.
One should first notice that the top-right point of each curve corresponds
to $\xi=0$, since both currents are maximum there.
Decreasing $\mu_{\LL}$ from zero slightly, the charge current
in the clogged region increases more sensitively than the energy current.
With decreasing $\mu_{\LL}$ further, the left part ($\xi<0$)
of the curve nearly overlaps with the right part ($\xi>0$).
Therefore, in a wide region including $\xi=0$, the ratio of the two currents
is nearly constant, and this is reminiscent of Wiedemann-Franz law
in thermal equilibrium.
The value of this ratio will be discussed in more detail later.  
This region is wide in the $\xi >0$ part for all $\beta_{\LL}$ and $\mu_{\LL}$.
The size of the $\xi<0$ part depends on the initial conditions.
It is wide at low temperatures (large $\beta_{\LL}$), but at higher temperatures
it shrinks quickly near half filling $\mu_{\LL}$ as the clogged behavior appears.  
This result reflects that the contributions of bound states decrease in both $j_{n}$ and $j_{e}$, and that of scattering states becomes dominant. Moreover, close to the two light cones $V_{\LL}$ and $V_{\RR}$, ratios of $j_{n}$ and $j_{e}$ are almost independent of $\beta_{\LL}$ and $\mu_{\LL}$. In this regime, $j_{n}$ and $j_{e}$ satisfy the relation $j_{e}\approx -4j_{n}$. This result is understood as follows. Since $\dddRR{v}{0}(k)=2\sin{k}$, the fastest quasiparticle, which reaches $\xi=V_{\RR}$, has a charge momentum $k=\pi/2$. Its bare energy is $e_{0}\paren{k=\pi/2}=-2u=-4$, therefore, $j_{e}\approx -4j_{n}$ close to $V_{\RR}$. For large $|\mu_{L}|$, $\dddLL{v}{0}(k)\approx 2\sin{k}$ as shown by Eq.~\eqref{eq:infdv1}. Therefore, close to $V_{\LL}$, the quasiparticles with charge momentum $k\approx-\pi/2$ mainly flow. Since $e_{0}\paren{k=-\pi/2}=-2u=-4$, the two currents also satisfy the relation $j_{e}\approx -4j_{n}$.

\subsection{Time dependence and stationary currents}

Let us now discuss the time evolution of physical quantities 
such as $n(x,t)$ and $j_n (x,t)$.  
For the partitioning protocol, they depend only on 
the ray $\xi = x/t$.  
For any fixed position $x$, physical quantities 
$O(x,t)$ approach their value at $\xi =0$ as time goes to infinity
\begin{equation}
O (x,t)  \stackrel{t \rightarrow \infty}{\longrightarrow} 
 O (\xi =0) + O'  \, \frac{x}{t} 
 + \frac{1}{2} O'' \, \frac{x^2}{t^2} + \cdots , 
\end{equation}
where $O'$ and $O''$ are the first- and second-order 
derivative of $O(\xi )$, respectively, at $\xi =0$.  
Thus, $O(\xi =0)$ is the stationary value at $t \rightarrow \infty$ 
at any position $x$. 
It is important that $O(x,t)$ approaches this value 
algebraically in time $t$, which implies that 
the system has no time scale characterizing its evolution 
in the long-time asymptotic region.  

The leading order correction is smaller for currents. 
For any conserved quantity such as particle number, energy, and 
magnetization, its local density $A(x,t)$ is related to the corresponding 
current $j_A (x,t)$ via the continuity equation, 
$d A(x,t) /dt + d j_A (x,t) / dx =0$. 
In the ray representation, it reads as 
\begin{equation}
  \xi A^{\prime} (\xi ) = j_A^{\, \prime} (\xi )
\label{eq:cont}
\end{equation}
and this concludes $j_A^{\, \prime} (\xi =0) =0$ 
unless $A'$ diverges accidentally at $\xi =0$.  
This proves that the current amplitude is local extremum 
at the origin $\xi =0$.  
Actually, in all the data of our calculations, 
the current amplitude is always maximum at $\xi =0$.  
This extremity also implies that the leading correction starts 
from the second order for various currents 
such as charge current, energy current, and spin current 
\begin{equation}
j_A (x,t)  \stackrel{t \rightarrow \infty}{\longrightarrow} 
 j_A (\xi =0) 
 + \frac{1}{2} A' \, \frac{x^2}{t^2} + \cdots , 
\end{equation}
where the identity $j_A '' = A'$ is used.  
Thus, currents converge to their stationary value 
faster than conserved quantities.  
\begin{figure}[tb]
\centering
\includegraphics[width=\columnwidth,clip]{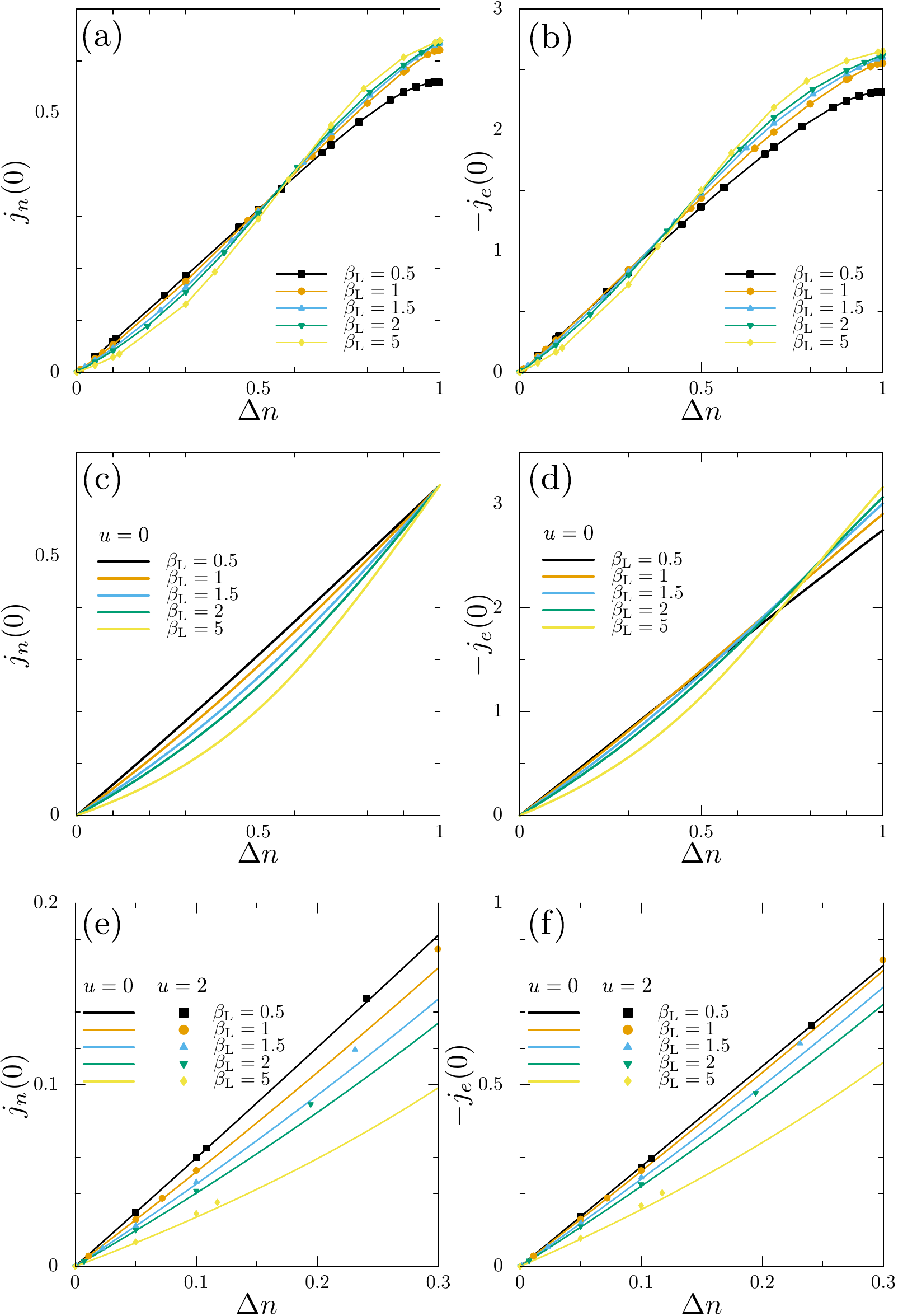}
\caption{
Stationary values of two currents, 
(a), (c), and (e) $j_n (\xi=0 )$ and 
(b), (d), and (f) $j_e (\xi =0)$, 
plotted as a function of $\Delta n = n^{\LL} - n^{\RR} = n^{\LL}$. 
(a) and (b) are the numerical data for the Hubbard model with $u=2$.  
(c) and (d) are the results of the non-interacting systems ($u=0$) 
calculated by the GHD theory.  
Note that the half filling case ($\Delta n =1$) is special, 
and $j_{n}(0)=2/\pi$ independent of $\beta_{\LL}$.   
(e) and (f) show the comparison in the small $\Delta n$ range 
with the non-interacting systems. 
Symbols show the data for $u=2$, while lines are the results 
for $u=0$. 
}
\label{fig:sne}
\end{figure}

We have calculated the dependence of stationary currents 
on initial conditions. 
One control parameter of the conditions
is the density difference between the left and right initial states
$\Delta n\equiv n^{\mathrm{L}}-n^{\mathrm{R}}=n^{\mathrm{L}}$.  
Another control parameter is the initial temperature 
in the left part $\beta^{-1}_{\LL}$.  
Fig.~\ref{fig:sne} (a) and (b) show the amplitude of 
the stationary particle density current $j_{n}\paren{\xi=0}$ and 
stationary energy current $j_{e}\paren{\xi=0}$, respectively, 
as a function of $\Delta n$ 
for $\beta_{\LL}=0.5,1,1.5,2,$ and $5$.  
In the small $\Delta n$ region, both currents increase linearly 
with $\Delta n$, and their slopes increase with increasing temperature $\beta^{-1}_{\LL}$. 
On the other hand, in the large $\Delta n$ region, temperature dependence reverses, and 
both currents decrease with increasing temperature.
We note that the temperature dependence of $j_{e}(0)$ turns out more complicated as we will show later.  

It is useful to compare these results with those for the non-interacting 
limit $u=0$.
For  $u=0$, the initial state is then characterized by 
the Fermi-Dirac distribution, 
and an electron with momentum $k$ propagates at velocity $v(k)=2\sin{k}$.
Its bare energy should be set as $e(k)=-2\cos{k}-2u$ corresponding to that for the interacting case $e_{0}(k)$. 
The stationary particle density current $j_{n}(0)$ is shown 
in Fig.~\ref{fig:sne}~(c) as a function of $\Delta n$.  
In contrast to the result for $u=2$ shown in Fig.~\ref{fig:sne}~(a), 
the current increases with increasing temperature 
in the whole region of $0 < \Delta n < 1$.
This result implies that the Coulomb interaction reverses  
the temperature dependence for $0.6 < \Delta n \le 1$.
The stationary energy current $j_{e}(0)$ is shown 
in Fig.~\ref{fig:sne}~(d). The temperature dependence reverses in both the interacting and non-interacting cases. 
Fig.~\ref{fig:sne}~(e)~and~(f) shows the comparison between 
the interacting and non-interacting cases in the small $\Delta n$ region. 
They agree very well for small $\Delta n$, and 
this demonstrates that Coulomb interaction has almost no 
effect in systems of dilute electrons, 
which is consistent with general understanding about correlation effects. 
This is because electrons can circumvent on-site repulsion 
effectively by their correlated motions. 
It is worth mentioning that the stationary particle density current for $u=0$
has a universal value independent of temperature at $\Delta n=1$. 
We can prove this analytically and its value is 
\begin{equation}
  j_n (0) = 2/\pi, \quad (u=0, \, \Delta n =1). 
\end{equation}  
Furthermore, the slope in the small $\Delta n$ region is asymptotically
\begin{align}
  \lim_{\Delta n \to 0}\frac{j_{n}(0)}{\Delta n}
  &=
  \frac{\sinh\paren{2\beta_{\LL}}}{\pi \beta_{\LL}~I_{0}\paren{2\beta_{\LL}}},
\label{eq3:jnratio}  
  \\
  \lim_{\Delta n \to 0}\frac{j_{e}(0)}{\Delta n}
  &=
  \frac{\paren{1-2u\beta_{\LL}}\sinh\paren{2\beta_{\LL}}
    -2\beta_{\LL}\cosh\paren{2\beta_{\LL}}}{\pi \beta^{2}_{\LL}~I_{0}\paren{2\beta_{\LL}}},
\label{eq3:jeratio}  
\end{align}
with $I_{0}$ being the zeroth order modified Bessel function of the first kind.
The slope of $j_{n}(0)$ monotonically increases with increasing $\beta^{-1}_{\LL}$, which is consistent with the behavior in Fig.~\ref{fig:sne}~(a).
On the other hand, the slope of $j_{e}(0)$
shows a nonmonotonic behavior with temperature $\beta^{-1}_{\LL}$.

\begin{figure}[tb]
\centering
\includegraphics[width=5.0cm,clip]{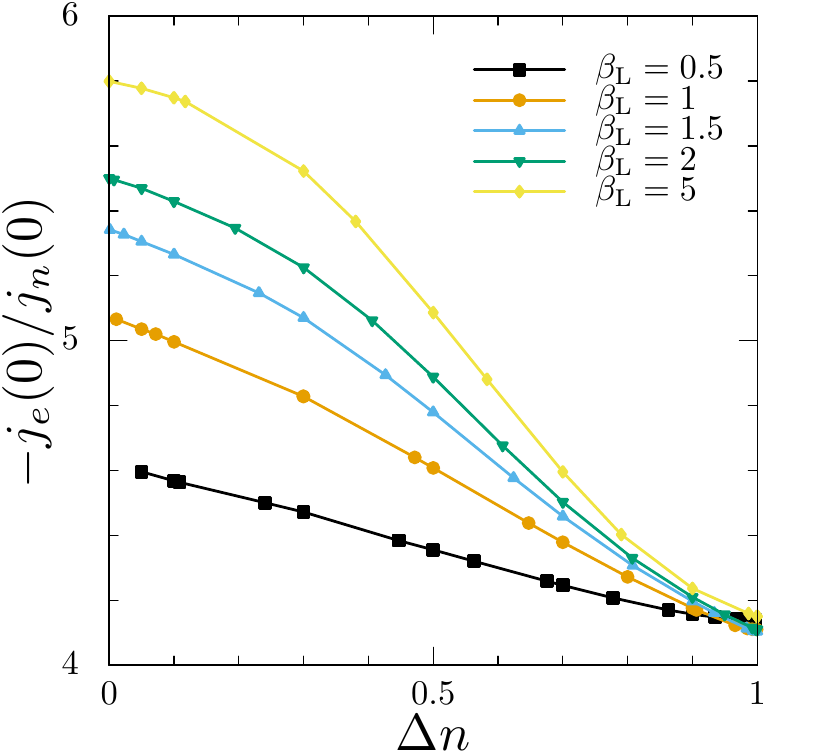}
\caption{
  Chemical potential dependence of ratio of the two stationary currents $j_{e}(0)$ and $j_{n}(0)$.
Five curves correspond to different temperatures $0.5\leq \beta_{\LL}\leq 5$.
}
\label{fig:rajnje}
\end{figure}

Finally, let us analyze the ratio of the two stationary currents $j_n (0)$ and $j_e (0)$
by examining its dependence on the initial conditions.
Wiedemann-Franz law in thermal equilibrium states that the ratio of
thermal conductivity to electric conductivity 
is proportional to temperature $\beta^{-1}$ aside from
a constant called Lorenz number\cite{test}.  
Therefore, if one source drives both charge and heat currents simultaneously,
one may expect that the ratio of two currents also follows
a similar scaling.
Since local thermodynamic potential is not well defined in nonequilibrium
state, we consider energy current instead of heat current.

Figure \ref{fig:rajnje} shows the ratio of the stationary
value of energy current $j_e (0)$ and particle density current $j_n (0)$,
which is equivalent to charge current.
The ratio is calculated from the data in Fig.~\ref{fig:sne} (a) and (b) and
plotted against $\Delta n = n^{\LL}$ for five temperature sets.
An important point is that the ratio depends on not only $\beta_{\LL}$
but also $\Delta n$, and this means that the proportionality relation
between the two currents is not so simple as Wiedemann-Franz law.
With approaching $\Delta n=0$, the ratio grows and energy current is
relatively enhanced than particle density current.
Another important feature is that the ratio has different temperature 
dependence
upon varying $\Delta n$.
For smaller values of $\Delta n$, the ratio is large and
shows large enhancement with lowering temperature. 
This is opposite to Wiedemann-Franz law.  
The ratio in the $\Delta n \rightarrow 0$ limit can be analytically 
evaluated using Eqs.~\eqref{eq3:jnratio} and \eqref{eq3:jeratio}:  
\begin{equation} 
  \lim_{\Delta n \to 0}\frac{j_{e}(0)}{j_{n}(0)}
  = \beta^{-1}_{\LL}- 2 \bigl[ u+\coth( 2\beta_{\LL} ) \bigr].
\end{equation}
This agrees very well with the limiting values in Fig.~11.
Therefore, this large temperature dependence is the one 
expected for noninteracting electrons in nonequilibrium state.
With increasing $\Delta n$, the ratio decreases.  
At the same time, the temperature dependence is
suppressed and very small near the half filling $\Delta n=1$.
In the very close vicinity, the temperature dependence reverses
and the ratio is suppressed with lowering temperature.
Therefore, strongly correlated electrons in the Hubbard model
show a nice proportionality of the two currents in a wide range
of spatio-temporal points, but their ratio is strongly suppressed 
by electron correlations, which grow near $\Delta n =1$.
The temperature dependence is even reversed at $\Delta n \approx 1$. 
These are an interesting characteristic feature of nonequilibrium
dynamics in correlated systems.

\section{Conclusions}
\label{sec:6}
In this paper, we have studied a nonequilibrium dynamics of the 1D Hubbard model based on GHD for the partitioning protocol.
We have analyzed initial conditions dependence of the profiles of densities of local conserved quantities and their currents.
In particular, we have found and analyzed the emergence of the clogged region, which has no charge current while energy current flows, when the initial 
left part is half-filled. First, we presented the analytical results in the case that the left initial state is at infinite temperature. We derived the conditions that quasiparticles' contributions to densities satisfy Eqs.~\eqref{eq:n_xi}~and~\eqref{eq:m_xi} and showed the existence of the clogged region analytically. We also showed general relationships between charge and spin currents Eq.~\eqref{eq:jnandjm} in the case of $\beta_{\LL}\mu_{\LL}\neq0$ and $\beta_{\LL}B_{\LL}\neq 0$ in the left part. Charge current is proportional to spin current, and their ratio is determined by the initial left conditions. This is a characteristic phenomenon of nested integrable systems, which have multiple degrees of freedom.

We have also numerically solved GHD equations to study charge and energy currents at finite temperatures. The initial state is set that the right part has no electron and the left part has no magnetic field.
We showed that, even at finite temperatures, the clogged region exists if $\mu_{\LL}=0$, and its width shrinks with decreasing the temperature. In the clogged region, the charge current carried by scattering states (i.e., real-$k$) is nearly canceled by counterflow carried by bound states (i.e., $k$-$\Lambda$ strings). On the other hand, the clogged region vanishes if $\mu_{\LL}\neq 0$ due to the change of the initial left particle density from unity. In this case, the contribution of scattering states dominates charge and energy currents.

Finally, we have analyzed initial conditions dependence of stationary charge and energy currents. The stationary currents were compared with the values in non-interacting case. For small $\Delta n$, where the Coulomb interaction has almost no effect, their results agree very well. In the non-interacting case, for all $0<\Delta n<1$, the particle density current grows always with increasing temperature, therefore, the reversed temperature dependence in the interacting cases is due to the Coulomb interaction.
The ratio of energy current to particle density current (equivalent to charge current)
is examined by varying the initial conditions.
For $\Delta n$ not so close 1, the ratio grows with decreasing temperature and its
temperature dependence converges to the formula (67) in the limit of $\Delta n \rightarrow 0$.
With increasing $\Delta n$, the temperature dependence becomes strongly
suppressed and we expect that this is due to the enhancement of electron correlation effects.
At $\Delta n \approx 1$, the temperature dependence is even reversed.
This suppression is an important characteristic of nonequilibrium dynamics
in strongly correlated electrons.

\section*{Acknowledgements}
We would like to thank Xenophon Zotos for useful discussions
as well as his kind teaching of the generalized hydrodynamic theory.  
Calculations in this work have been partly performed using
the facilities of the Supercomputer Center at ISSP,
the University of Tokyo.

\appendix
\section{Filling functions in a thermal equilibrium}
\label{app:filling}

In this Appendix, we explain how to calculate the filling functions 
$\bce{\vartheta_{a}}$ for a thermal equilibrium 
state parametrized by 
inverse temperature $\beta$, chemical potential $\mu$, 
and magnetic field $B$. 
It is convenient to use $\{ \eta_a \}$ defined in Eq.~\eqref{eq2:eta}, 
which are equivalent to the filling functions.  
They are determined as the solutions of the TBA equations~\cite{10.1143/PTP.47.69}:
\begin{equation}
  \log \eta_{a} (\clam ) 
  =
  s \star \log \bigl[ (1+\eta_{a-1}) (1+\eta_{a+1}) \bigr] 
  \Big|_{\clam},
\label{eq:tba4}
\end{equation}
for $|a|\geq 2$ and 
\begin{align}
   \log \eta_{-1} (\clam ) 
   &=  s \star \log \bigl( 1 + \eta_{-2} \bigr) \big|_{\clam}
    - s~\hat{\star}~ 
      \left[  \cos k \cdot \log \bigl( 1+\eta_{0} \bigr) 
      \right] \Big|_{\clam},
\label{eq:tba1}
\\
  \log \eta_{0} (k) 
  &=-2\beta 
   \left[ \cos k + 2s * \Re \sqrt{ 1-(\clam-\imi u)^2 } \ \Big|_{k}
   \right]
\nonumber\\
  &\phantom{=} +
  s* \log \bigl[ \bigl( 1+\eta_{-1} \bigr) \big/ 
                 \bigl( 1+\eta_{1} \bigr) \bigr] \Big|_{k},
\label{eq:tba2}
\\ 
  \log \eta_{1} (\clam ) 
  &= s \star \log \bigl( 1+\eta_{2} \bigr) \Big|_{\clam}
   - s~ \hat{\star} 
     \bigl[ \cos k \,  \log \bigl( 1+\eta_{0}^{-1} \bigr) \bigr] 
    \Big|_{\clam},
\label{eq:tba3}
\end{align}
with the boundary values
\begin{equation}
   \lim_{a \to -\infty}
   \frac{\log \eta_{a} (\clam ) }{a}
   =2\beta \mu,
\quad 
   \lim_{a \to\infty}
   \frac{\log \eta_{a} (\clam ) }{a}
   =2\beta B.
\label{eq:tba5}
\end{equation}
Here, three types of convolutions are used 
\begin{align}
s \star f \big|_{\clam}
&\equiv 
\int_{-\infty}^{\infty} d \clam ' \,  
s (\clam - \clam ') \, f ( \clam ' ),
\label{app1:sstar}
\\
s \, \hat{\star} \, f \big|_{\clam}
&\equiv 
\int_{-\pi}^{\pi} d k \ 
s (\clam-\sin k ) \, f (k) , 
\label{app1:hatstar}
\\
s * f \big|_{k}
&\equiv 
\int_{-\infty}^{\infty} d \clam  \, 
s ( \sin k - \clam ) \, f (\clam ), 
\label{app1:aster}
\end{align}
with the kernel function
\begin{equation}
s (\clam ) \equiv 
\frac{1}{4u} 
\sech  \frac{\pi \clam }{ 2u }.
\label{eq:s}
\end{equation}
Note that only the part of $\eta_0$ has an inhomogenous term 
in the above infinite set of coupled integral equations.

\section{Calculation of dressed quantities}
\label{app:dr}

We explain here how to calculate the derivative of 
dressed momentum $\{ \drst{k}{a} \}$ 
and dressed energy $\{ \drst{e}{a} \}$ 
for a given set of the filling functions $\{\vartheta_a \}$.  
In this Appendix, we drop the argument $\xi$ 
such as $\drst{k}{a} (w,\xi ) \rightarrow  \drst{k}{a} (w) $ etc., 
since $\{ \drst{k}{a} \}$ and $\{ \drst{e}{a} \}$ 
at different $\xi$'s are solved independently.  
At each $\xi$, 
$\{ \drst{k}{a} \}$ and $\{ \drst{e}{a} \}$ 
follow the linear integral equations 
with kernels including the filling functions.  
Since their kernels are common, 
it is convenient to use the following vector
\begin{equation}
 \bm{X}_a (w) = 
\left[ 
\begin{array}{c}
\drk{a} (w) \\[4pt] 
\dre{a} (w)
\end{array}
\right] . 
\end{equation} 

With this, the integral equations read 
\begin{equation}
\bm{X}_a (\clam )
=
\left. 
s \star 
\left( \bar{\vartheta}_{a-1} \bm{X}_{a-1}{}
      +\bar{\vartheta}_{a+1} \bm{X}_{a+1}{}
\right) \right|_{\clam}, 
\quad (|a| \ge 2)
\label{eq:de4}
\end{equation}
and 
\begin{align}
\bm{X}_{-1} {(\clam )}
&=
\left. 
s \star 
\left( \bar{\vartheta}_{-2} \bm{X}_{-2} \right)
\right|_{\clam}
-
\left. 
s~\hat{\star}~ 
\left( \bar{\vartheta}_{0} \bm{X}_{0}  \right)
\right|_{\clam},
\label{eq:de1}
\\
\bm{X}_{0} (k) 
&=
\left[ 
\begin{array}{c}
 1 \\[2pt] 
2 \sin k
\end{array}
\right]
- \cos k \cdot 
\left. 
s* 
\left[ 
\begin{array}{c}
{k}^{\prime}_{-1} \\[2pt] 
{e}^{\prime}_{-1}  
\end{array}
\right]
\right|_{k}
\nonumber\\
&\phantom{=}
- \cos k \cdot 
\left. 
s * 
\bigl( \bar{\vartheta}_{1}  \bm{X}_{1}
      -\bar{\vartheta}_{-1} \bm{X}_{-1} \bigr)
\right|_{k},
\label{eq:de2}
\\
\bm{X}_{1}(\clam ) 
&=
s \star 
\bigl( \bar{\vartheta}_{2} \bm{X}_{2} \bigr) \big|_{\clam}
+
s~ \hat{\star}~ 
\bigl( \vartheta_{0} \bm{X}_{0} \bigr) \big|_{\clam},
\label{eq:de3}
\end{align}
with the inhomogeneous term determined by 
the bare values 
\begin{align}
\left[
\begin{array}{c}
{k}^{\prime}_{-1} (\clam) \\[2pt]
{e}^{\prime}_{-1} (\clam)
\end{array}
\right] 
&= 
\Re \left\{ 
\frac{-2}{\sqrt{1- (\clam - iu)^2 }} 
\left[
\begin{array}{c}
 1 \\[2pt] 
2 (\clam - iu )
\end{array}
\right] 
\right\} . 
\end{align}
Recall $\bar{\vartheta}_a = 1 - \vartheta_a$.  
The boundary conditions are 
$
   \lim_{a \to \pm \infty}
   {a}^{-1} \, \bm{X}_{a} (\clam ) =0
$.
Note that the above vector representation is used 
only for shortening the equations.  
$\{ \drk{a} \}$ and $\{ \dre{a} \}$ 
are decoupled, and 
they are solved separately.

\section{Lower bound of dressed velocities}
\label{app:dv}

In this Appendix, we prove the lower bound of the dressed velocities in Eq.~\eqref{eq:dvine}. 
It is sufficient to prove for $a>0$. 
Equations ~\eqref{eq:infdv2} leads
\begin{equation}
\dddLL{v}{a} (\clam)+ 2 
= 
2 \, \frac{ P_{a} - P_{a+2} }{ Q_{a} - Q_{a+2} },
\label{eqA:vdr}
\end{equation}
with 
\begin{align}
&\left[ 
\begin{array}{c}
 P_a \\ Q_a 
\end{array}
\right] 
\equiv 
\Re 
\left\{ 
\frac{ f_{a-1}^{-1} }{ \sqrt{1- (\clam - iau)^2} } 
\left[ 
\begin{array}{c}
1 + \clam - iau \\ 1 
\end{array}
\right] 
\right\}
\nonumber\\
&=
\frac{u \sinh |x|}{a^{-1}\sinh a|x|} 
\int_{-\pi}^{\pi} \frac{dk}{2\pi} 
\frac{ 1 }{ a^2 u^2 + (\clam - \sin k)^2 }
\left[ 
\begin{array}{c}
1 + \sin k \\ 1 
\end{array}
\right],
\label{eqA:PQ}
\end{align}
where $x = \bar{B}$.
The two factors in the denominator, 
$a^{-1} \sinh a|x|$ and 
$a^2 u^2 + (\clam - \sin k )^2$, both increase
monotonically with increasing $a >0$.  
If $x=0$, $u \sinh |x| / (a^{-1} \sinh a|x|) = u$ and 
this does not change the following argument.  
The integrand is positive definite and semidefinite, respectively, 
for $Q_a$ and $P_a$. 
Combining these shows that $P_a$ and $Q_a$ decrease monotonically 
with increasing $a$. 
This proves the positivity of the RHS of Eq.~\eqref{eqA:vdr}. 

For $a<0$, Eq.~\eqref{eqA:PQ} holds again 
with the only change $x=\bar{\mu}$, 
and therefore $P_a$ and $Q_a$ decrease monotonically with 
increasing $|a|$. 
Thus, Eq.~\eqref{eq:dvine} is proved
\begin{equation}
 -2 < \dddLL{v}{a} (\clam ) , 
 \quad ({}^\forall a \ne 0).
\end{equation}

\section{Total distributions in the clogged region}
\label{app:rhotp}

We here prove Eq.~\eqref{eq4:rhotp} in the limit $\beta_{\LL}=0$
for the Fourier transformation 
of the total distributions $\{ \tilde{\rho}_a^t \}$ 
in the clogged region $V_{\LL} \le \xi \le \xi_{-\infty}^{-}$.  
In this region, the filling functions of 
long $k$-$\Lambda$ strings do not change from their values 
in the initial equilibrium state in the left part; 
see Eq.~\eqref{eq:inffill}.  
Namely, there exist a negative integer $a_{*}$ such that 
$\vartheta_a (\clam , \xi) = \vartheta_a^{\LL} = f_a^{-2}$ 
for all $a < a_{*} < 0$.  
Note that $\vartheta_a^{\LL} (\clam )$ is independent of $\clam$ 
in the infinite-temperature limit as shown in 
Eqs.~\eqref{eq:inffill}-\eqref{eq:fa}.  

The total distributions follow Takahashi's equations~\cite{10.1143/PTP.47.69} at each $\xi$:
\begin{equation}
\rho^{t}_{a}\paren{\clam}
= \left. 
s \star 
\bigl( \bar{\vartheta}_{a-1}\rho^{t}_{a-1}
      +\bar{\vartheta}_{a+1}\rho^{t}_{a+1} \bigr)
\right|_{\clam},
\label{eqD:te1}
\end{equation}
for $|a|\geq 2$ and
\begin{align}
\rho^{t}_{-1}\paren{\clam}
=&\asub{s\star\paren{\bar{\vartheta}_{-2}\rho^{t}_{-2}}}{\clam}
+\asub{s~\hat{\star}\paren{\bar{\vartheta}_{0}\rho^{t}_{0}}}{\clam},
\label{eqD:te2}\\
\rho^{t}_{0}\paren{k}
=&\frac{1}{2\pi}-\frac{1}{2\pi}\cos{k}\cdot\asub{s*k^{\prime}_{-1}}{k}
\nonumber\\
&-\cos{k}\cdot\asub{s*\paren{\bar{\vartheta}_{1}\rho^{t}_{1}
+\bar{\vartheta}_{-1}\rho^{t}_{-1}}}{k},
\label{eqD:te3}\\
\rho^{t}_{1}\paren{\clam}
=&\asub{s\star\paren{\bar{\vartheta}_{2}\rho^{t}_{2}}}{\clam}
+\asub{s~\hat{\star}\paren{\vartheta_{0}\rho^{t}_{0}}}{\clam},
\label{eqD:te4}
\end{align}
where $\bar{\vartheta}_{a}=1-\vartheta_{a}$, and the argument $\xi$ is dropped. The convolutions are defined in Eqs.~\eqref{app1:sstar}-\eqref{app1:aster}.
The boundary conditions are $\lim_{a\to \pm \infty}a^{-1}\rho^{t}_{a}\paren{\clam}=0$.

The Fourier transform of the total fillings 
$\{ \tilde{\rho}_a^t (p , \xi )\}$ satisfy 
the following recurrence equation 
for $a+1<a_{*}$ 
\begin{equation}
   2 \cosh up \cdot\, 
   \tilde{\rho}^{t}_{a} (p, \xi )
 = \bar{\vartheta}^{\LL}_{a+1} 
   \tilde{\rho}^{t}_{a+1} (p, \xi )
  +\bar{\vartheta}^{\LL}_{a-1}
   \tilde{\rho}^{t}_{a-1} (p, \xi ) ,
\label{eq:rec1}
\end{equation}
where we have used the Fourier transform of $s(\clam )$
\begin{equation}
 \tilde{s} (p) \equiv 
\int_{-\infty}^{\infty} d\clam \, 
e^{-ip\clam} s(\clam ) 
= {\textstyle \frac{1}{2}} \sech up .
\label{eqD:sp} 
\end{equation}
Its general solution for $p\neq 0$ is given by~\cite{10.1143/PTP.46.401,10.1143/PTP.47.69}
\begin{align}
   \tilde{\rho}^{t}_{a} (p, \xi )
  &= A_{1} (p, \xi ) f_{a}
   \left( 
       \frac{\, e^{a    u|p|} }{ f_{a+1} }
      -\frac{   e^{(a-2)u|p|} }{ f_{a-1} }
   \right)
\nonumber
\\
  &+A_{2} (p, \xi ) f_{a} 
   \left( 
       \frac{ e^{-a    u|p|} }{ f_{a+1} }
      -\frac{ e^{-(a-2)u|p|} }{ f_{a-1} }
   \right).
\label{eq:recsol1}
\end{align}
The total distribution $\tilde{\rho}^{t}_{a}\paren{p}$ 
should not diverge as $a \to -\infty$, 
and this imposes the condition $A_{2}\paren{\clam,\xi}=0$. 

The case of $p=0$ is special. The two terms in Eq.~\eqref{eq:recsol1}
are not linearly independent, 
and another linearly independent solution exists. 
We have proved that this additional solution 
also diverges as $a \to -\infty$, 
and its proof is given in Appendix \ref{app:rec}. 
Thus,  $\tilde{\rho}^{t}_{a}\paren{p,\xi}$ 
can be represented for any $p$ as
\begin{equation}
  \tilde{\rho}^{t}_{a} \paren{p,\xi}
  = A_{1} (p, \xi) f_{a}
   \left( 
       \frac{\, e^{a    u|p|} }{ f_{a+1} }
      -\frac{   e^{(a-2)u|p|} }{ f_{a-1} }
   \right) . 
\label{eq:recsol2}
\end{equation}
The coefficient $A_{1}$ is to be determined from Takahashi's equations~\eqref{eqD:te1}-\eqref{eqD:te4}. From Eq.~\eqref{eq:vartheta}, for $a\geq a_{*}$, $\bar{\vartheta}_a (\clam , \xi)$ is either $\bar{\vartheta}^{\RR}_a (\clam)$ or $\bar{\vartheta}^{\LL}_a (\clam)$, and therefore $A_{1}$ depends on the initial state in the right part through $\bar{\vartheta}_a (\clam , \xi)$ for $a\geq a_{*}$.

\section{Divergent solution of the recurrence relation}
\label{app:rec}

In this Appendix, we consider the recurrence relation 
\eqref{eq:rec1} for $p=0$
\begin{equation}
   2 r_{a-1} 
=
 \bar{\vartheta}^{\LL}_{a}   r_{a} 
+\bar{\vartheta}^{\LL}_{a-2} r_{a-2}, 
\quad (a < a_{*} < 0),
\label{eq:rec2}
\end{equation}
and will prove that it has a solution which diverges as $a \to -\infty$. 
Here we use a shorthand notation 
$r_{a} = \tilde{\rho}^{t}_{a} (p=0, \xi )$ for simplicity, 
and note that $r_a \ge 0$ for any $a$.  
We drop the argument $\xi$, since different $\xi$-component 
do not couple in the relation above. 
As defined in Eqs.~\eqref{eq:inffill}-\eqref{eq:fa},  
$\bar{\vartheta}^{\mathrm{L}}_{a}$=
$\sinh^2 \bar{\mu} / \sinh^2 [(a-1) \bar{\mu}]$. 
The recurrence relates consecutive three terms, and 
the initial two terms determine the whole series.  
The relation is homogeneous, i.e., 
if $\{ r_a \}$ is a solution, $\{ \lambda r_a \}$ is also a solution 
for any $\lambda$.  
Thus, $r_a$ is a linear function of two initial values 
and may be represented as follows
\begin{align}
 r_a 
= 
  \bigl(             r_{a_{*}-1} 
       + \alpha   \, r_{a_{*}-2} \bigr) \, A_a 
+ \bigl( \beta   \, r_{a_{*}-1} 
       +            r_{a_{*}-2} \bigr) \, B_a , 
\label{eqB:sol}
\end{align}
where $\alpha$, $\beta$, $A_a$ and $B_a$ 
are independent of the initial values.   

We will show that one of the two terms, say $B_a $ part, 
diverges as $a \to -\infty$. 
To prove this, it suffices 
to show divergence for some set of the initial values, 
and we choose the case of 
$0 < r_{a_{*}-1} < r_{a_{*}-2}$. 
For this case, 
we rewrite the recurrence relation \eqref{eq:rec2} 
into the following form and show that 
the difference of neighboring terms increases 
monotonically with increasing $|a|$
\begin{align}
r_{a-2} - r_{a-1}
&=
 \frac{ 2 - \bar{\vartheta}^{\LL}_{a-2}}%
      {     \bar{\vartheta}^{\LL}_{a-2} } \, 
      r_{a-1}
-\frac{ \bar{\vartheta}^{\LL}_{a}   }%
      { \bar{\vartheta}^{\LL}_{a-2} } \, 
      r_{a}
\nonumber\\
&> r_{a-1} - r_{a}.
\end{align}
Here, we have used the relation 
$0 < \bar{\vartheta}^{\LL}_{a} < \bar{\vartheta}^{\LL}_{a-2} < 1$ 
for $a<0$.  

This proves that $\{ r_a \}$ starting from general initial values 
diverges as $a \to -\infty$ 
at least linearly, generally faster 
\begin{align}
 \lim_{a \to -\infty} r_a = \infty, \quad 
 \lim_{a \to -\infty} \frac{r_a}{|a|} 
  =  {}^{\exists} C > 0, \mbox{ or } \infty.  
\end{align}
Since $r_a$ is the total distribution of $k$-$\clam$ strings,  
this should not diverge as $a \to -\infty$. 
This imposes a constraint for the two initial values 
discussed in Appendix~\ref{app:rhotp}.  
Their ratio should be properly chosen such that the coefficient 
of the divergent part $B_a$ vanishes in Eq.~\eqref{eqB:sol}.

\section{Energy current in the clogged region}
\label{app:iene}
In this Appendix, we show an example where energy current is nonvanishing in the clogged region. Let us consider the case of $|\bar{\mu}|=\bar{B}$ in the limit $\beta_{\LL}=0$ and a restricted clogged region $V_{\LL}<\xi<V_{\LL,1}$. In this case, $\rho^{t}_{0}$ and $\protect\accentset{\circ}{v}_{0}$ are independent of $\xi$, and energy current can be analyzed further. We explicitly show that energy current flows at least for $u>1$ and $\vartheta^{\RR}_{0}\paren{k}<1/2$.

\subsection{Uniformity of $\rho^{t}_{0}$ and 
$\protect\accentset{\circ}{v}_{0}$}
\label{app:rt}
The restricted clogged region $V_{\LL}<\xi<V_{\LL,1}$ is 
special in that 
the distribution $\rho_0^t (k, \xi )$ and 
the dressed velocity $\protect\accentset{\circ}{v}_{0} (k, \xi)$ 
are uniform and do not vary with $\xi$ in its inside.  
This property is important for discussing 
$\xi$-dependence of energy density, and 
let us prove this first.  

In the restricted clogged region, there exists the integers $a_{*}=a_{\star}=0$.
Therefore, 
the total distributions of strings for $a<0$ 
have the form in Eq.~\eqref{eq:recsol2} 
in Appendix.~\ref{app:rhotp}.  
The result for $a>0$ is similarly written as 
\begin{align}
\tilde{\rho}^{t}_{a} (p,\xi )
  = A_{2} (p,\xi ) f_{a} 
\left(
 \frac{e^{-ua |p|}}{f_{a-1}}
-\frac{e^{-u(a+2)|p|}}{f_{a+1}}
\right) . 
\label{eqF:apm}
\end{align}
These coefficients $A_{1}$ and $A_{2}$ 
are to be determined from 
Takahashi's equations~\eqref{eqD:te1}-\eqref{eqD:te4}.
Adding the Fourier transforms of Eqs.~\eqref{eqD:te2}~and~\eqref{eqD:te4} 
and recalling $\bar{\vartheta}_a$'s are constant, 
one obtains 
\begin{align}
\sum_{a= \pm 1}
  &\tilde{\rho}_{a}^t (p, \xi ) 
- 
\tilde{s} (p)
\sum_{a= \pm 2}
  \bar{\vartheta}_{a} \, 
  \tilde{\rho}_{a}^t (p, \xi ) 
\nonumber\\
&= \int d\clam \, e^{-i p \clam} \, 
\int dk \, s (\clam - \sin k ) \rho_0^t (k , \xi )
\label{eqF:rho+1-1A}
\\
&= \tilde{s} (p) \int dk \, e^{-i \sin k} \rho_0^t (k , \xi )
=  \tilde{s} (p) J_0 (p) , 
\label{eqF:rho+1-1B}
\end{align}
where $J_n (p)$ is the Bessel function of order $n$
and $\tilde{s}(p)$ is the Fourier transform of 
$s(\clam )$ given in Eq.~\eqref{eqD:sp}.  
The $k$-integral in Eq.~\eqref{eqF:rho+1-1B} has been 
calculated using Eq.~\eqref{eqD:te4} 
and all the terms except $(2\pi )^{-1}$ 
have null contribution owing to the identity 
\begin{equation}
 \int_{-\pi}^{\pi} dk \, f (\sin k ) \, \cos k = 0 .  
\label{eqF:identity}
\end{equation}
One should note that the $\xi$-dependence has disappeared.  
We have also calculated the LHS of Eq.~\eqref{eqF:rho+1-1A} with 
evaluating factors $f$'s by Eq.~\eqref{eq:fa} and found that 
the result is 
$\tilde{s} (p) [ f_1 A_2 (p,\xi ) + f_{-1} A_1 (p,\xi ) ]$. 
Thus, we have obtained the following relation
\begin{align}
  f_{1}  A_{2} (p,\xi )  +f_{-1} A_{1} (p,\xi ) 
=J_{0}\paren{p} . 
\end{align}
This relation holds in the whole restricted clogged region, 
irrespective the values of $\bar{\mu}$ and $\bar{B}$.  

In the special case of $|\bar{\mu}|=\bar{B}$, 
we can derive a simple expression of $\rho_0^t$ 
using the obtained relation.  
In this case, the two $f$ factors coincide $f_{a} = f_{-a}$ 
$({}^\forall a \ne 0)$, 
and this also leads to $\bar{\vartheta}_{a}=\bar{\vartheta}_{-a}$. 
This simplifies Eq.~\eqref{eqD:te3} as 
\begin{align}
2\pi \, &\rho^{t}_{0} (k,\xi ) 
=
1 - \cos k \cdot s*k^{\prime}_{-1} \big|_{k}
\nonumber\\
&-\bar{\vartheta}^{\LL}_{-1} \cos k 
\int dp \, e^{- i p \sin k} \, 
\tilde{s} (p) J_0 (p) 
\left(  e^{-u|p|} - \frac{ e^{-3u|p|} }{f_{-2}}  \right)
\nonumber\\
&=1 - 
\bar{\vartheta}^{\LL}_{-1} \cos k \cdot k^{\prime}_{-2} (\sin k )
= 2\pi \, \rho^{t}_{0}  (k) , 
\hspace{0.3cm}   (\xi < V_{\LL , 1}). 
\label{eqF:rtind}
\end{align}
Therefore, $\rho^{t}_{0}$ does not depend on $\xi$ in the restricted 
clogged region. 
Namely, it is uniform and unchanged from 
the distribution in the initial equilibrium state in the left part.

Let us also examine the dressed velocity $\accentset{\circ}{e}$ 
in the restricted clogged region.  
We can prove that the derivative of dressed energy 
$\dre{0}$ is independent of $\xi$ similarly.  
The Fourier transform of Eq.~\eqref{eq:de4} is written as 
\begin{align}
  \tilde{\accentset{\circ}{e}}^{\, \prime}_{\pm a} (p,\xi )
  =B_{\pm} (p,\xi ) f_{\pm a}
  \left( 
  \frac{ e^{-a     u|p|} }{ f_{\pm (a-1)} }
 -\frac{ e^{-(a+2) u|p|} }{ f_{\pm (a+1)} } \right) , 
\label{eqF:bpm}
\end{align}
for $a\geq 1$. 
Performing calculations similar to those for 
Eqs.~\eqref{eqF:rho+1-1A}-\eqref{eqF:rho+1-1B}, one obtains 
\begin{align}
 f_{1} B_{+} (p,\xi ) - f_{-1} B_{-} (p,\xi ) 
&=\int dk \, e^{-\imi p \sin k} \, \dre{0} (k)
\nonumber\\
&=-4 \pi i \, J_{1} (p).  
\end{align}
Evaluating Eq.~\eqref{eq:de2} with this result 
and repeating manipulation similar to those in Eq.~\eqref{eqF:rtind}, 
we finally obtain 
\begin{equation}
\dre{0} (k,\xi ) 
= 
2 \sin{k} - \vartheta^{\LL}_{-1}\cos k\cdot e^{\prime}_{-2} (\sin k ) , 
\hspace{0.3cm} (\xi < V_{\LL , 1}).
\end{equation}
Thus, $\dre{0}$ is also independent of $\xi$ in the restricted 
clogged region when $|\bar{\mu}|=\bar{B}$. 

With these results obtained, it is straightforward to 
calculate the dressed velocity $\ddd{v}{0}=\dre{0}/\drk{0}$. 
Since $\drk{0} = 2\pi \rho^{t}_{0}$, both of $\dre{0}$ and $\drk{0}$ 
are independent of $\xi$, and this guarantees the uniformity 
of dressed velocity
\begin{align}
\ddd{v}{0} (k, \xi )
= \frac{ \displaystyle
2 \sin{k} - \vartheta^{\LL}_{-1}\cos k\cdot e^{\prime}_{-2} (\sin k )  }
{ \displaystyle 
 1         - \vartheta^{\LL}_{-1}\cos k\cdot k^{\prime}_{-2} (\sin k )  }, 
\hspace{0.5cm} 
(\xi < V_{\LL , 1}).
\label{eq:iene3}
\end{align}

\subsection{Analysis of energy current}
\label{app:exe}
One can prove that the energy density can be 
expressed by $\rho_{0}$ and $\rho^{h}_{-1}$ alone. 
Using Eq.\eqref{eq:e} and $\rho=\rho^t - \rho^h$, 
we decompose this into three components 
\begin{equation}
e (\xi ) = 
\int dk \, e_0 (k) \rho_0 (k) 
+ I_1 - I_2 , 
\end{equation}
with
\begin{equation}
\left(
\begin{array}{c}
I_1  \\[4pt] I_2
\end{array}
\right)
\equiv 
\sum_{a<0} \int d\clam \, e_a (\clam ) 
\left(
\begin{array}{c}
\rho_a^t (\clam , \xi) \\[4pt]
\rho_a^h (\clam , \xi) 
\end{array}
\right) . 
\end{equation}
The energy functions satisfy the recurrence relation 
\begin{equation}
 e_a (\clam ) = s \star (e_{a-1} + e_{a+1} ) \big|_\clam,
\end{equation}
and using this for $I_2$, one obtains 
\begin{align}
I_2 
&= \int d\clam 
\left[ 
\sum_{a \le -1} s \star e_{a+1} \Big|_\clam 
\rho_a^h 
+ 
\sum_{a \le -1} s \star e_{a-1 } \Big|_\clam 
\rho_a^h 
\right].
\end{align}
For $I_1$, let us use Eqs.~\eqref{eqD:te1}-\eqref{eqD:te4} 
and rewrite this term as
\begin{align}
I_1 
= \int d\clam 
\left[ 
\sum_{a \le -2} e_{a+1} s \star \rho_a^h \Big|_\clam 
+
\sum_{a \le -1} e_{a-1} s \star \rho_a^h \Big|_\clam 
+
e_{-1} s \hat{\star} \rho_0^h \Big|_\clam 
\right] . 
\end{align}
Notice the identity 
$\int d \clam \, f \, (s \star g )
= \int d \clam  \, (s \star f) \,  g $, 
and one finds that 
each summation in $I_1$ cancels the corresponding 
part in $I_2$ except the term of $a=-1$ in the first summation.  
Thus, the difference of the two components is given as 
\begin{equation}
 I_1 - I_2 
= \int d\clam 
\left[ 
 s \hat{\star} e_{-1} \big|_\clam 
(\rho_0^t - \rho_0 )
- 
 s \star e_{0} \big|_\clam 
\rho_{-1}^h 
\right],
\end{equation}
where $\rho_0^h$ has been replaced by $\rho_0^t - \rho_0$.  
To evaluate the term including $\rho_0^t$ in this, 
we use Eq.~\eqref{eqD:te3} and find that the result does not 
depend on $\xi$
\begin{align}
&\Delta I_{12}^0 
\equiv
\int \,d\clam \, 
e_{-1}  \, s \, \hat{\star} \, \rho_0^t \, \Big|_\clam 
\nonumber\\
&= 
\int d\clam dk \, 
e_{-1} (\clam )
s (\clam - \sin k) \rho_0^t (k)
= 
\int \frac{dk}{2\pi} \, 
s * e_{-1} \big|_k . 
\end{align}
Namely, among the terms for $\rho_0^t$ in Eq.~\eqref{eqD:te3}, 
only the term $1/(2\pi )$ survives in the last expression of 
$\Delta I_{12}^0$.  
To derive this, we have used the identity Eq.~\eqref{eqF:identity} 
for $f(\sin k) = s (\clam - \sin k) s (\sin k - \clam ')$.  
Combining these results, the energy density is 
represented by $\rho_0$ and $\rho_{-1}^h$ alone: 
\begin{align}
e(\xi )
=&\Delta I_{12}^0 
+\int d k 
\Bigl[ e_{0} (k)-  s*e_{-1} \big|_k \Bigr]
\rho_{0} (k, \xi ) 
\nonumber\\
&-
\int d\clam 
\Bigl[ e_{-1} (\clam ) - s \star e_{-2} \big|_{\clam} \Bigr]
\rho^{h}_{-1} (\clam,\xi ) , 
\label{eqF:eng}
\end{align}
where $s \star e_0$ has been replaced by 
$e_{-1} - s \star e_{-2}$.  
This result holds for any $\xi$, not limited to 
the restricted clogged region.  

In the restricted clogged region $V_{\LL} < \xi < V_{\LL , 1}$, 
$\rho^{h}_{-1}$ is determined by $\rho^{h}_{0}$ through Eq.~\eqref{eqF:apm} 
and we can further simplify the result \eqref{eqF:eng}  
and the Fourier transform of Eq.~\eqref{eqD:te2} : $f_{-1}A^{-}\paren{p,\xi}=\int \diff k e^{-\imi p \sin{k}}\rho^{h}_{0}\paren{k,\xi}$.
Therefore, using $f^{2}_{-1}=f_{-2}+1$ and $\tilde{e}_{a}\paren{p}=4\pi J_{1}\paren{p}e^{a u |p|}/p$,
\begin{align}
&\int\diff\clam\bck{e_{-1}\paren{\clam}-\asub{s\star e_{-2}}{\clam}}\rho^{h}_{-1}\paren{\clam,\xi}
\nonumber\\
=&\int\frac{\diff p}{2\pi}e^{\imi p \clam}\frac{4\pi J_{1}\paren{p}}{p}\paren{e^{-u|p|}-\frac{e^{-2u|p|}}{2\cosh{u p}}}
\nonumber\\
&\times \bar{\vartheta}^{\LL}_{-1}\int \diff k e^{-\imi p \sin{k}}\rho^{h}_{0}\paren{k,\xi}\paren{e^{-u|p|}-\frac{e^{-3u|p|}}{f_{-2}}}
\nonumber\\
=&\int \diff k \asub{s*\paren{\bar{\vartheta}^{\LL}_{-1}e_{-1}-\vartheta^{\LL}_{-1}e_{-3}}}{k}\rho^{h}_{0}\paren{k,\xi}
\nonumber\\
=&\int \diff k \asub{s*\paren{\bar{\vartheta}^{\LL}_{-1}e_{-1}-\vartheta^{\LL}_{-1}e_{-3}}}{k}\bck{\frac{1}{2\pi}-\vartheta_{0}\paren{k,\xi}\rho^{t}_{0}\paren{k}}.
\end{align}

Thus, the energy current is represented as 
\begin{align}
&e\paren{\xi}
=\vartheta^{\LL}_{-1}\int \frac{\diff k}{2\pi} 
\asub{s*\paren{e_{-1}+e_{-3}}}{k}
\nonumber\\
&+\int \diff k 
\bck{e_{0}\paren{k}-\vartheta^{\LL}_{-1}\asub{s*\paren{e_{-1}+e_{-3}}}{k}}\vartheta_{0}\paren{k,\xi}\rho^{t}_{0}\paren{k}
\nonumber\\
=&\vartheta^{\LL}_{-1}\int \frac{\diff k}{2\pi}e_{-2}\paren{\sin{k}}
\nonumber\\
&+\int \diff k 
\bck{e_{0}\paren{k}-\vartheta^{\LL}_{-1}e_{-2}\paren{\sin{k}}}\vartheta_{0}\paren{k,\xi}\rho^{t}_{0}\paren{k}.
\label{eqF:iene}
\end{align}
This differs from the initial equilibrium value 
in the left part by the following quantity
\begin{align}
 &e\paren{\xi}-e^{\LL}\nonumber\\
 =&\int d k 
\Bigl[ e_{0} (k) -\vartheta^{\LL}_{-1}e_{-2} (\sin k ) \Bigr]
\Bigl[ \vartheta_{0} (k,\xi ) -\vartheta^{\LL}_{0}  \Bigr]
    \rho^{t}_{0}\paren{k}.
\label{eq:iene2}
\end{align}
Thus, the $\xi$-dependence in the energy density comes 
from the part of $\vartheta_{0} - \vartheta^{\LL}_{0}$.  

Finally, let us quantify the spatial variation $e' (\xi)$ further.  
Since $\ddd{v}{0}$ does not depend on $\xi$ in the $\xi$-region,
differentiating Eq.~\eqref{eq:vartheta} leads to
\begin{align}
\partial_{\xi}\vartheta_{0} (k,\xi ) 
=
\delta \left( \xi-\dddLL{v}{0} \paren{k} \right) \, 
\left[ \vartheta^{\RR}_{0} (k) - {\textstyle \frac12} \right],
\end{align}
where we have used $\vartheta^{\LL}_{0}=\frac{1}{2}$ for $|\bar{\mu}|=\bar{B}$.
Note that the equation $\dddLL{v}{0}\paren{k}=\xi$ 
has two solutions $K_{-}(\xi ) < K_{+}(\xi )$ for 
$V_{\LL}=\min_{k}\dddLL{v}{0}\paren{k} < \xi < \max_{k}\dddLL{v}{0}$.  
Therefore, the spatial variation of energy density 
$e^{\prime}(\xi )$ is given as
\begin{align}
e^{\prime} (\xi )
=&
\sum_{\alpha=\pm}
\Bigl[ 
  e_{0} \bigl( K_{\alpha} \bigr) - 
  \vartheta^{\LL}_{-1}	
   e_{-2} \bigl( \sin K_{\alpha} \bigr) 
\Bigr]
  \rho^{t}_{0} \bigl( K_{\alpha} \bigr)
\bigl|  K^{\prime}_{\alpha} (\xi) \bigr| 
\nonumber\\
  &\hspace{0.3cm}
\times 
\left[ 
  \vartheta^{\RR}_{0} \bigl( K_{\alpha} \bigr) 
  - {\textstyle \frac12 }
\right]
\Big|_{K_{\alpha} = K_{\alpha} (\xi)}
, 
\label{eq:dedxi}
\end{align}
where the prime denote the differentiation by $\xi$, 
and we have used the relation 
$1=
\frac{d \, \dddLL{v}{0}}{d\, k}
k^{\prime}(\xi )$.
The dependence of the parameters 
in the right part $\paren{\beta_{\RR},\mu_{\RR},B_{\RR}}$ 
comes only from the part of $\vartheta^{\RR}_{0}$.  
This result clearly shows that 
$e^{\prime}(\xi )$ is generally nonzero, 
except for accidental or the trivial case such as 
$\vartheta^{\RR}_{0} (k) =\frac12$ for all $k$ 
as in the left equilibrium.
In particular,  in the case of $u>1$ and 
$\vartheta^{\RR}_{0} (k) < \frac12$, 
the two terms represented with bracket
in the sum are both negative, and 
thus $e^{\prime}(\xi ) > 0$ for all $\xi$ in this region.

\bibliographystyle{unsrt}


\end{document}